\documentclass[aps,prx,onecolumn,superscriptaddress,noshowpacs]{revtex4}

\usepackage[intlimits]{amsmath}
\usepackage{amsfonts}
\usepackage{graphics}
\usepackage{subfigure}
\usepackage[usenames]{color}
\usepackage{epstopdf,epsfig}
\usepackage{color}
\usepackage{indentfirst}

\usepackage{graphicx}
\usepackage{amsmath}
\usepackage{amsfonts}

\begin{document}

\title{Explicit Hamiltonians Inducing Volume Law for Entanglement 
Entropy in Fermionic Lattices}

\author{Giacomo Gori}
\affiliation{CNR-IOM DEMOCRITOS Simulation Center, Via Bonomea 265, 
I-34136 Trieste, Italy}

\author{Simone Paganelli}
\affiliation{Dipartimento di Scienze Fisiche e Chimiche, Universit\`{a} dell'Aquila, via Vetoio, I-67010 Coppito-L'Aquila, Italy}
\affiliation{International Institute of Physics, Universidade Federal
do Rio Grande do Norte, 59078-400 Natal-RN, Brazil}

\author{Auditya Sharma}
\affiliation{International Institute of Physics, Universidade Federal
do Rio Grande do Norte, 59078-400 Natal-RN, Brazil}
\affiliation{School of Chemistry, The Sackler Faculty of Exact Sciences, 
Tel Aviv University, Tel Aviv 69978, Israel}

\author{Pasquale Sodano}
\affiliation{International Institute of Physics, Universidade Federal
do Rio Grande do Norte, 59078-400 Natal-RN, Brazil}
\affiliation{Departemento de Fis\'ica Teorica e Experimental, 
Universidade Federal do Rio Grande do Norte, 59072-970 Natal-RN, Brazil}
\affiliation{INFN, Sezione di Perugia, Via A. Pascoli, I-06123, Perugia,
Italy} 

\author{Andrea Trombettoni}
\affiliation{CNR-IOM DEMOCRITOS Simulation Center, Via Bonomea 265, 
I-34136 Trieste, Italy}
\affiliation{SISSA and INFN, Sezione di Trieste, Via Bonomea 265, I-34136 
Trieste, Italy}


\begin{abstract}
\noindent 

We show how the area law for the entanglement entropy may be violated by free fermions on 
a lattice and look for conditions leading to the emergence of a volume law. We give an 
explicit construction 
of the states 
with maximal entanglement entropy based on the fact that, once a bipartition of the lattice 
in two complementary sets $A$ and $\bar{A}$ is given, the states with maximal entanglement 
entropy (volume law) may be factored into Bell-pairs (BP) formed by two states with support on $A$ and $\bar{A}$.
We then exhibit, for translational invariant fermionic systems on a lattice, 
an Hamiltonian whose ground state is such to yield an exact volume law. As expected, the corresponding 
Fermi surface has a fractal topology. We also provide some examples 
of fermionic models for which the ground state may have an entanglement entropy $S_A$ between 
the area and the volume law, building an explicit example of a one-dimensional free fermion model 
where $S_A(L) \propto L^\beta$ with $\beta$ being intermediate between 
$\beta=0$ (area law) and $\beta=1$ (BP-state inducing volume law). For this model, the 
dispersion relation has a ``zig-zag'' structure leading to a fractal Fermi surface whose 
counting box dimension equals, for large lattices, $\beta$.
Our analysis clearly relates the violation of the area law for the entanglement entropy of the ground state to the 
emergence of a non-trivial topology of the Fermi surface. 
\end{abstract}

\maketitle

\section{Introduction}\label{intro}
The study of entanglement in quantum systems has been
a major 
field of research in the last two decades. Motivations for this 
are various and important since 
entanglement provides not only a characterization of quantum states 
\cite{amico08,horodecki09,guhne09} and a pathway on how 
to simulate them with numerical tools such as DMRG \cite{schollwock05}  and tensor network states \cite{TN}, but 
it helps also to characterize quantum phase 
transitions \cite{osterloh02,osborne02,gu04} 
and to detect novel quantum phases, 
including topological phases. Non-trivial, explicit examples of the 
use of entanglement-related quantities, such as entanglement 
spectrum and negativity, range from quantum Hall states 
\cite{li08, thomale10, thomale10_bis} to 
Bose-Hubbard \cite{alba13} and Kondo models \cite{bayat12,bayat14}. 
In addition the study of entanglement allows to characterize 
the computational power of quantum phases 
\cite{cui12,dechiara12,lepori13,franchini13}.

A key quantity entering the characterization of entanglement
is provided by the Entanglement Entropy (EE). For its 
definition one takes a partition of a given system in two subsystems 
$A$ and $\bar{A}$ (the complement of $A$), 
determines the reduced density matrix 
of a subsystem 
(say, of $A$) $\rho_A$ 
by tracing out the degrees of freedom in $\bar{A}$ and then computes its entropy: 
$S_A=-Tr_{A} ( \rho_A \ln{\rho_A})$ \cite{horodecki09}. The 
celebrated {\em area law} \cite{srednicki93,eisert10} for the EE refers 
to the fact that typically the EE grows 
as the boundary of the subsystem $A$: i.e., for a system in $d$ dimensions 
and a subsystem of size $L$ having volume $\sim L^d$ and area $\sim L^{d-1}$,   
$S \sim L^{d-1}$ 
according of the area law \cite{srednicki93,callan94,calabrese04}. 

The study of EE in various models 
has been recently a subject of intense research: EE can be explicitly 
computed in non-interacting systems of bosons and fermions 
\cite{peschel03,cramer06,wolf06,gioev06,peschel09,calabrese12}, 
including trapped fermions 
\cite{vicari12},  in integrable 
\cite{cardy08,its09, its09a} 
and 1D critical models \cite{ rico03,korepin04,calabrese05,calabrese09a}, and in spin chains 
with long-range interactions \cite{koffel12}. 
An important result is that, for gapless 
$1d$ integrable systems, the EE grows as $\ln{L}$, and that the prefactor is 
proportional to the central charge of the model. The next-leading 
term of the EE has been studied as well: e.g., 
for $2d$ systems the size-independent 
constant entering $S$ is the so-called topological EE \cite{kitaev06,levin06}. 
Furthermore, the EE of a subsystem made of two disjoint intervals has been 
also intensively studied \cite{furusaki_pasquier,calabrese09,igloi10}: 
for these issues we refer the interested readers to \cite{jphysa09}.

A first possible deviation from the area law is provided 
by logarithmic corrections: as shown in 
\cite{wolf06,gioev06,swingle10,ding12}, 
for (critical) fermionic systems of dimension $d$, the EE of a subsystem of 
size $L$ typically grows 
as $S_A\sim L^{d-1} \ln{L}$ (this result does not hold for bosonic systems 
\cite{cramer06}). An explicit expression for the prefactor 
entering $S_A$ in dimension larger than $1$ may be given using the 
Widom conjecture \cite{gioev06}, and is found in remarkable agreement 
with numerical results \cite{barthel06,li06}. 
Entropy bounds for reduced density matrices of fermionic states 
were given in \cite{klich06,carlen14}; 
the role of disorder was also investigated \cite{refael04}, and it was shown 
that the momentum space entanglement spectrum reveals the location of 
delocalized states in the energy spectrum \cite{mondragon13} and that 
the entanglement structure depends only on the probability 
distribution of the length of the effective bonds \cite{ramirez14}. 
Furthermore, allowing for long-range interactions leads to 
a logarithmically 
diverging EE in gapped noncritical models \cite{eisert06}, 
in spin chains \cite{ding08} and Bose-Einstein condensates 
\cite{ding09}. Non-local exponentially decaying 
couplings were considered in \cite{shiba13}: at intermediate distances a 
volume law is observed, but as soon as $L$ becomes of 
the order of the length scale of the decay of the couplings the area law is 
recovered.

Pertinent inhomogeneous couplings 
in simple spin chain Hamiltonians with 
only nearest-neighbor interactions 
have been shown to induce a volume law in absence 
of translational invariance \cite{vitagliano10}.
A violation of the area law for bosonic systems with Bose 
surfaces was analyzed in \cite{lai13}. 
Fermi liquids are expected to obey the area law while non-Fermi 
liquids in $2d$, although  have been shown to satisfy the area law, are at 
the border between area law and non-area law EE 
\cite{metlitski10,swingle13}. A construction 
of a translation invariant fermionic state 
violating the area law 
was explicitly given in \cite{huijse13}, when the Fermi surface is a  
Cantor-like set. Non-logarthmic deviations from the area law were also 
observed in \cite{pouranvari14}, where two different kinds of disordered 
fermionic chains were considered. An  analysis 
of EE in spin chains having long-range interactions and a fractal 
Fermi surface for the associated Jordan-Wigner fermions has been presented 
in \cite{fannes03,farkas05}: in particular in \cite{farkas05} it was shown that 
EE for all translational-invariant pure states is at least logarithmic and there is 
an arbitrary fast sublinear entropy growth.

States supporting an area law for the EE are not maximally entangled 
and yet maximally entangled states 
have been deeply studied both for their intrinsic 
interest and in connection to
quantum information protocols:   
maximally multiqubit entangled states up to eight qubits in qubit registers 
were reported in \cite{brown05,borras07,tapiador09,zha11,zha12}, while 
applications of absolutely maximally entangled states (i.e., 
multipartite quantum states maximally entangled with respect to 
any possible bipartition) to a variety of quantum information protocols, 
including  quantum secret sharing schemes 
\cite{helwig12,helwig13_bis} and open-destination teleportation protocols 
\cite{helwig13}.

In this paper we show that a volume law for the EE of the ground state may emerge 
in fermionic lattices: to avoid ambiguities, 
we say that the  a violation of the area law is obtained if, 
apart from logarithmic corrections,
the EE scales as $S_A\sim L^\beta$, with $\beta>d-1$. In particular, for 
$1d$ chains, a violation of the area law  corresponds to $\beta>0$ and 
for $\beta=1$ we have the volume law.
To set the notation, we write the Hamiltonian of (spinless) free 
fermions hopping on a generic lattice as
\begin{equation}
H=-\sum_{I,J} c^\dag_I t_{IJ} c_J.
\label{Hamiltonian}
\end{equation}
The lattice has $N_S$ sites and
its connectivity is characterized by  the hopping matrix $t_{IJ}$ with $t_{IJ}=t^*_{JI}$; of course $c_I$ 
and $c_I^\dagger$ are the annihilation and creation fermionic operators on the site $I$. The number of fermions is $N_T$, and the filling is then $f=N_T/N_S$
($0\le f \le 1$).  
The sites of the lattice are denoted by capital letters 
$I,J=1,\cdots,N_S$ while the sites of a generic subsystem having $L$ sites are denoted  by (small) letters 
$i,j=1,\cdots,L$. If the system 
is in the pure state $\vert \Psi \rangle$, the EE of the subsystem $A$ is given by 
\begin{equation}
S_{A}=-\sum_{\gamma=1}^{L} \left[ \left( 1-C_\gamma \right)
\ln{\left( 1-C_\gamma \right)} + C_\gamma \ln{C_\gamma} \right],
\label{res}
\end{equation}
where  $C_\gamma$ is one of the $L$ eigenvalues of the correlation matrix 
\begin{equation}
C_{ij} =\langle \Psi \vert c_i^\dag c_j \vert \Psi \rangle.
\label{correl_intr}
\end{equation} 
In Appendix \ref{app:A}, following \cite{peschel09}, we provide an explicit expression of the 
correlation matrix $C_{ij}$ if  $\vert \Psi \rangle$ is the ground-state 
of the Hamiltonian \eqref{Hamiltonian}, which is the situation we are going to mostly consider.

From Eq. (\ref{res}), one easily sees that for the EE to satisfy the volume law, one may construct 
a state for which each $C_\gamma$ is equal to $1/2$ since for this state $S_A=-L \ln \frac{1}{2}$. In the 
following, we shall provide a method for the construction of such states and give some examples of 
Hamiltonians supporting as a ground state a state with maximal EE.
We get the remarkable result that, associated with these states, Fermi surfaces with nontrivial 
topology naturally emerge.

At first sight one might think that a volume law could emerge only as a result of introducing 
a long-range hopping matrix $t_{IJ}$ in the Hamiltonian (\ref{Hamiltonian}). Our analyses shows 
that this is not the case since, at least for translational invariant systems, it is rather the 
topology of the Fermi surface which really matters, as also pointed out in previous analysis \cite{swingle10}. 
Indeed we demonstrate that, give a partition of the single-particle Hilbert space to  orthogonal 
subspaces  $\mathcal{A}$ and $\bar{\mathcal{A}}$ the state yielding maximal EE may be factorized 
into Bell-pairs formed by  two states belonging to  $\mathcal{A}$ and $\bar{\mathcal{A}}$. We call 
such state Bell-paired state (BP-state). As we shall see explicitly, for translational-invariant 
Hamiltonians, the BP-states are highly  nonlocal in the space and the Fermi surfaces have nontrivial topology.

The paper is organized as follows: in Sec.  \ref{sec1} we analyze a free fermion model with 
hoppings decaying as a power law with exponent $\alpha$.  We find that the EE obeys to the area 
law for each finite and positive $\alpha$, even tough, for $\alpha<1$ the energy is not extensive 
\cite{campa09}; only for $\alpha=0$, i.e. for the fully connected lattice, one has a volume law for 
the EE since $S_A\propto L$. Unfortunately, the fully connected lattice is pathological in many 
respects, since the Fermi level is infinitely degenerate and 
the number of sites of a given 
subsystem $A$ is at the same time its volume and its surface 
(defining on a graph the volume of a subgraph as the number 
of vertices on it and the surface as the number of vertices on it 
linked to vertices outside the subgraph itself \cite{magan,hoory}). 

In Sec. \ref{topol} we explicitly construct, for any given lattice and arbitrary filling $f$, 
the general form of the states rendering the EE and all the R\'enyi entropies maximal. This construction allows for to have an explicit momentum representation of the state with maximal entanglement entropy.
In Sec. \ref{hamiltonians} we provide explicit examples of Hamiltonians supporting a BP-state 
as the ground state and analyze the topology of their Fermi surface.
Sec. \ref{concl} is devoted to our concluding remarks, and in the appendices  \ref{app:A} 
\ref{app:LRphase}  \ref{app:fcn} we provide the reader with technical details about the 
models described in the main text. In Appendix \ref{sec:random} we analyze the violation 
of the area law of the ground state for a free fermionic model which is not translational invariant.
 
\section{Free Fermions with non-local power-law hoppings} \label{sec1}
In this Section we focus our attention on a translationally invariant 
one-dimensional chain, 
with non-local hopping matrix $t_{I,J}$  given by 
\begin{equation}\label{tt}
t_{I,J} = \left\{
\begin{array}{ll}
0 & I=J,\\
\frac{t}{|I-J|_{p}^\alpha} & I \neq J,
\end{array} \right.
\end{equation}
where the distance $|\cdot|_{p}$, due to periodic boundary 
conditions, is defined as 
\begin{equation}
|I-J|_{p}=\min(|I-J|, N_S-|I-J|).
\label{definition_dist}
\end{equation} 

The hopping matrix, being translationally
invariant, is readily diagonalized: its eigenstates 
are given by plane waves and, in the thermodynamic limit, the energy 
spectrum is
\begin{equation}
\varepsilon_k = - 2t\ell_{\alpha}(k),
\end{equation}
where $k=2\pi n_k/N_S$ belongs to the first Brillouin zone 
($n_k=-N_S/2,\cdots,N_S/2-1$ for even 
$N_S$) and 
\begin{equation}
\ell_\alpha(k)= \sum_{m=1}^{\infty} \frac{\cos{(mk)}}{m^\alpha},
\label{elle_funct_app}
\end{equation}
(with $\alpha>1$). As usual, even  if
for $\alpha \le 1$ the ground-state energy in the 
thermodynamic limit diverges,
one can make the energy extensive by the so-called 
Kac rescaling \cite{campa09}. 

\begin{figure}[t]
\centerline{
{\includegraphics[width=1\columnwidth]{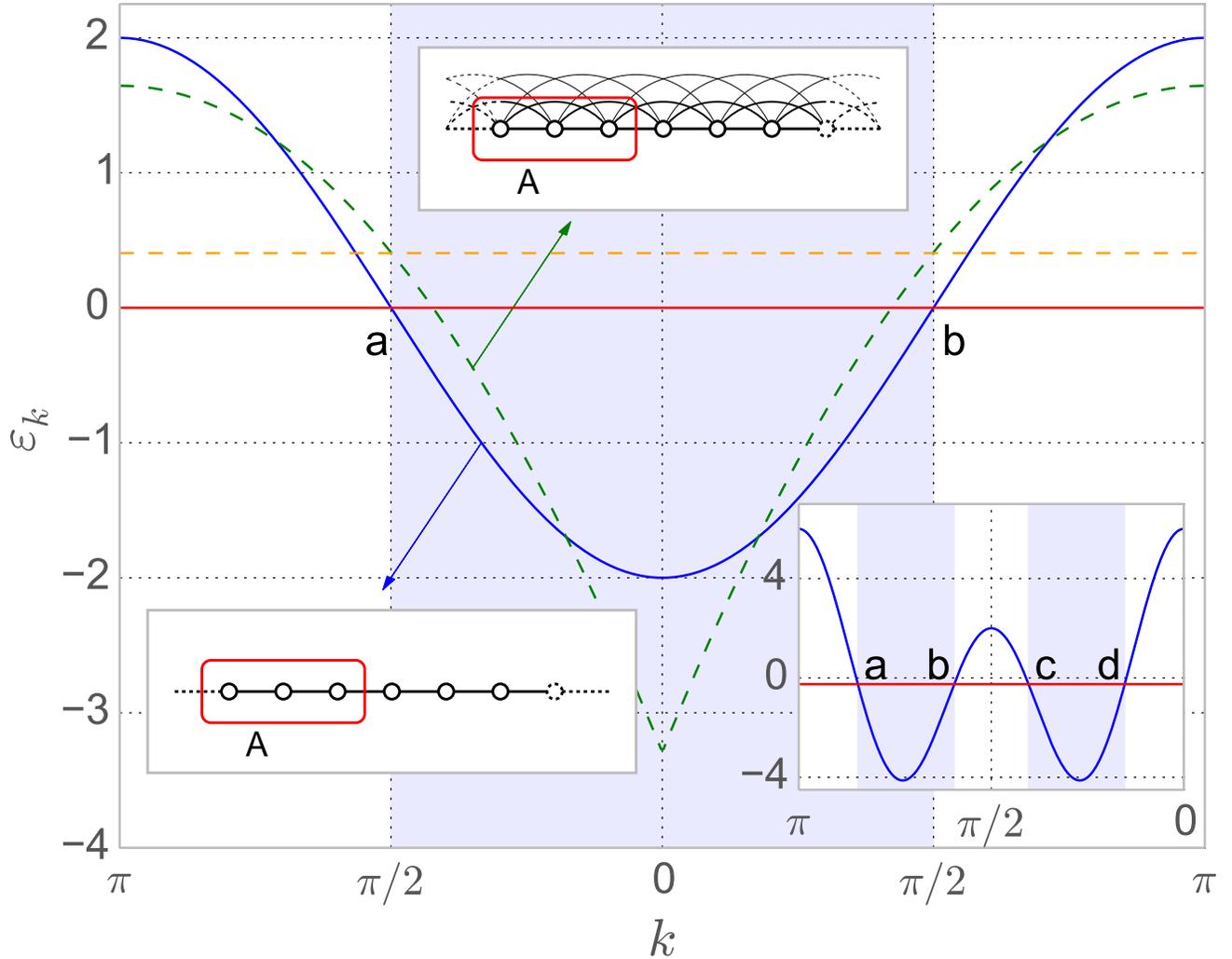}}
}
\caption{Energy spectra $\varepsilon_k$ (in units of $t$) of a model
with nearest-neighbor hopping corresponding 
to $\alpha \to \infty$ (solid line) 
and non-local hoppings with $\alpha=2$ (dashed line) 
in the thermodynamic limit. 
The horizontal lines represent the Fermi
energy at half-filling for both models.
The regions of $k$-space comprised within
the points $a$ and $b$ are the occupied
wave vectors, indicated in the whole diagram
with a (pale blue) shading. In the white (top and bottom left) 
the hopping structures of both models are represented - in the bottom right inset 
the energy spectrum for a next-nearest-neighbor model having 
four Fermi points (denoted by $a$, $b$, $c$ and $d$) is plotted ($t_1=t$, $t_2=-2t$).}
\label{fig1}
\end{figure}

The function $\varepsilon_k$ is plotted for two values
of $\alpha>1$ in Fig.\ref{fig1}. 
One sees that the spectra are monotonic for $k>0$ and $k<0$,
and, thus, the filling of the momentum eigenstates  
leading to the occupation of the Fermi sea is 
the same for each value of $\alpha>1$. 
The same result holds also for 
$0<\alpha \le 1$ for any finite number of sites (Appendix \ref{app:LRphase}).
As a consequence, the EE does not change 
since the correlation matrix (\ref{correl_intr}) 
depends only on the ground state (and not on the spectrum); in the thermodynamic 
limit, $S_A \sim \ln{L}$ for each $\alpha>0$ for any filling $f$, just as it happens 
if the hopping $t_{IJ}$ was short range \cite{wolf06}. A similar 
analysis, yielding the same results, may be carried out also for 
$t<0$ and $t = (-1)^{i-j}\vert t \vert$.

Things change for $\alpha=0$. 
Here, it is easy to verify that the single-particle energy spectrum is made of 
a non-degenerate ground-state and of an $(N_S-1)$-fold degenerate excited 
state, 
implying that the many-body ground-state is highly degenerate. 
In addition, the Fermi surface passes from a two-point set (as it happens for 
$\alpha>0$) 
to a continuous set. 
In Appendix \ref{app:fcn} we show that, for $N_S \gg 1$, 
\begin{equation}
S_{A}=-L\left[\left(1-f\right)\ln\left(1-f\right)+f\ln\left(f\right)\right];
\label{S_fully}
\end{equation}
in particular, $S_A=L \ln{2}$ for $f=1/2$. This corresponds to an 
equal {\em a priori} probability of occupation of all degenerate states 
by the available particles.

The fully connected hopping model does not have 
a specific dimensionality $d$: however, if we think it as the 
$\alpha \to 0$ limit of a $d$-dimensional long-range hopping model and 
$A$ is a cubic subsystem 
with size ${\cal L}$, then the number $L$ of sites of $A$ is given by 
$L \sim {\cal L}^d$. 

It appears as we already obtained a volume law for the EE. Unfortunately, for 
the  fully connected hopping model, the number of sites $L$ 
of the subsystem $A$ is at the same time the volume and the surface of $A$ 
in the sense that 
all the $L$ sites of $A$ are linked with the other sites of the rest of 
the system $\bar{A}$. In addition 
the mutual information between $A$ and $\bar{A}$ is vanishing for 
$N_S \gg 1$ and, thus, the emergence of a volume 
law corresponds here to nothing but the transition to a classical state.
 
The analysis carried out in this Section shows that, for translational 
invariant systems, long-range hoppings alone are 
not enough to guarantee the emergence of the volume law for the EE and it appears that 
 the structure of the Fermi 
surface is bound to play a key role in the behavior of the EE.
In particular, for a translational invariant chain and 
a quadratic Hamiltonian of the form \eqref{Hamiltonian},  
all the models having Fermi wave vectors 
$k_a$ and $k_b$ at the points $a,b$ 
of Fig.\ref{fig1} have the same correlation matrix, which, in the continuum
limit, reads
\begin{equation}
C_{ij}=\int_{k_a}^{k_b} \frac{dk}{2\pi} e^{ik(i-j)},
\label{Cij_AB}
\end{equation}
leading to the same EE (we remind 
that our Hamiltonian \eqref{Hamiltonian} does not 
include any ``superconducting'' $c_I^\dagger c_J^\dagger, 
c_I c_J$ terms, that would change the correlation matrix \eqref{Cij_AB}).
If $-k_a=k_b \equiv k_F$ then the well known result 
$C_{ij}=\sin{\left[k_F(i-j)\right]} / \left[ \pi \left( i-j \right)\right]$ 
of the nearest-neighbor free fermionic chain is recovered 
\cite{peschel03}.

To better clarify the role played by the Fermi surface let us  consider the energy spectrum represented 
in the bottom right inset of Fig.\ref{fig1} having $4$ 
Fermi wave vectors $k_a,k_b,k_c,k_d$ at the points $a,b,c,d$: the same argument
leading to \eqref{Cij_AB} yields
\begin{equation}
C_{ij}=\int_{k_a}^{k_d} \frac{dk}{2\pi} e^{ik(i-j)}-
\int_{k_b}^{k_c} \frac{dk}{2\pi} e^{ik(i-j)}.
\label{Cij_ABCD}
\end{equation}
(this formula can be easily generalized to Fermi surfaces 
with wave vectors $k_1,\cdots,k_{2m}$).
As a result, the EE depends only on $k_a,k_b,k_c,k_d$ and not on other details of the energy spectrum.

\section{States with maximal entanglement entropy}\label{topol}
\begin{figure}[t]
\centerline{
{\includegraphics[width=1\columnwidth]{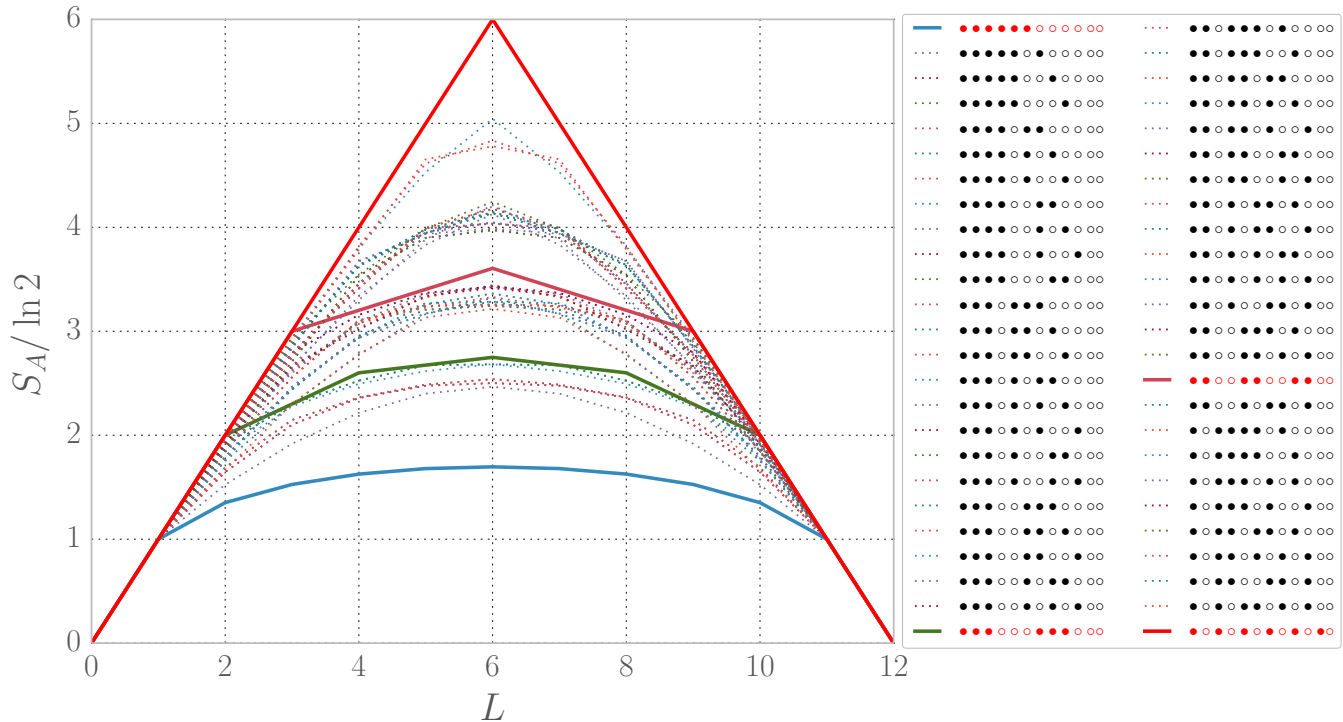}}
}
\caption{EE in terms of the subsystem 
length $L$ for all the combinatorially distinct
Fermi surfaces for a system of $N_S=12$ sites at
half-filling. In the right legend a filled (empty) dot 
denote a filled (unoccupied) momentum eigenstate. The states are ordered
with respect to increasing $k$ (modulo $2\pi$).
The states having a periodic structure in $k$-space
are highlighted in the legend with a color red
and are plotted in the left figure 
with thick continuous lines: the lower thick blue line refers 
to momenta occupied up to $k_F$ (resulting in logarithmic EE), 
the upper thick red line refers to the state 
with alternating filled momenta (having linear EE), the central thick scarlet 
and green  
lines to the states having 
sequence of two (three) filled momenta and two (three) holes.}
\label{fig2}
\end{figure}

In order to elucidate the role played by
 the Fermi surface for constructing a maximal entangled state 
in a translationally invariant
free fermionic lattice, it is instructive to look at \emph{all} the possible
Fermi surfaces arising in 1-dimensional systems of small size $N_S$.
This task is simplified if one  notices that
circular shifts ($k\rightarrow k+2 n_k \pi/N_S, n_k\in
\mathbb{Z}$) and reflections ($k\rightarrow -k$)
of the Fermi surface do not alter the EE.
As a result, the number of Fermi surfaces yielding different values of the EE is much reduced; from 
combinatorics this number is obtained by counting all the distinct reversible bracelets \cite{OEIS,riordan80}.
The result is shown  in Fig.\ref{fig2}
for a system with $N_S=12$ sites
at half-filling.
As one can see, the Fermi surfaces exhibiting
the maximal EE are realized with an alternating
filling of the wave vectors periodic in $k$-space with period $2$, while Fermi surfaces exhibiting higher periodicity in $k$-space,
have a piecewise linear behavior. The minimal
EE  for a system of $12$ sites at half filling is achieved with a Fermi surface
made up of two points, i.e. made occupying $6$ states with adjacent wave vectors.
Similar findings are obtained for different system sizes and fillings.
This procedure allows to explicitly 
construct a state of maximal EE for  small  fermionic systems through 
a  pertinent filling of the  Fermi sea. In addition, once the sistem size and the filling are given, it selects 
the Fermi surface for which a volume law emerges. 

In the following, we shall generalize 
the above result to systems of finite, but arbitrarily large, size $N_S$.

As shown in Sec.  \ref{intro}, the EE between $A$ and $\bar{A}$  for a non-interacting fermionic system in its 
ground-state is determined by the correlation matrix \eqref{correl_intr}, which may be usefully rewritten as
\begin{equation}\label{eqn:corrbis}
C_{i,j} = \sum_{b \in B} \langle i | b \rangle \langle b | j \rangle \quad 
\left( i, j \in A \right),
\end{equation}
where $B$ labels  the set of
single particle states entering in (\ref{eqn:corrbis}). 

To compute the EE one needs to find the eigenvalues $C_\gamma$ of the matrix $C_{i,j}$. In the following, 
we denote with $\mathcal{A}$ and $\mathcal{B}$  two nonorthogonal subspaces of the single-particle state 
space $\mathcal{H}$ such that
$\mathcal{A}=
\mathrm{span} \{ |i\rangle, i \in A\}$ and $\mathcal{B}=
\mathrm{span} \{ | b \rangle,  b \in B\}$. We then define $P_\mathcal{A}$ and $P_\mathcal{B}$ as the 
projection operators over the subspaces $\mathcal{A}$ and $\mathcal{B}$, respectively.
Upon introducing the operator

\begin{equation}
\Gamma= P_\mathcal{A} P_\mathcal{B} P_\mathcal{A},
\label{definiton_C}
\end{equation}
one has that

\begin{equation}
C_{i,j}=\langle i \vert \Gamma 
\vert j \rangle,
\end{equation}
with $i,j \in A$. 
As a result the EE can be written as
\begin{equation}\label{eqn:Sgamma}
S_A=S_{\mathcal{A},\mathcal{B}} = -\mathrm{Tr} \left[ \Gamma \ln{\Gamma} + 
\left( 1-\Gamma \right) \ln{\left( 1-\Gamma \right)} \right].
\end{equation}
With the notation used in (\ref{eqn:Sgamma}), the symmetries of $S_A$ are made manifest 
since  $S_{\mathcal{A},\mathcal{B}}= S_{\mathcal{A},\mathcal{\bar{B}}}$ 
and $S_{\mathcal{A},\mathcal{B}}= S_{\mathcal{\bar{A}},\mathcal{B}}$ (where $\mathcal{\bar{B}}$
is the orthogonal complement of $\mathcal{B}$). 

In the following we shall determine, for a given $\mathcal{A}$,  the vector space  $\mathcal{B}$ for 
which the  EE is maximal. As we shall see, for translational invariant free fermionic lattices, this 
amounts to determine the topology of the Fermi surface maximizing the EE.

The EE is strictly upper bounded 
by the dimension of
the smallest space between $\mathcal{A}$ and $\bar{\mathcal{A}}$ times $\ln 2$, since in 
Eq. (\ref{eqn:Sgamma}) natural logarithms have been used. 
Of course, $\dim \mathcal{A}=|A|$ and $\dim \bar{\mathcal{A}}=|\bar{A}|$, where $|A|=L$ ($|\bar{A}|=N_S-L$) 
is the cardinality of the set $A$ ($\bar{A}$). As a result 
\begin{equation} 
S_{\mathcal{A},\mathcal{B}} \leq S_\text{max} = \ln{2}  \cdot
\min(|A|,  |\bar{A}|) .
\end{equation} 
We shall now explicitly construct the states satisfying
this upper bound. We observe that,  in the construction of maximal EE states, 
we do not need to fix the dimension of $\mathcal{B}$, i.e. the number of fermions 
($\dim \mathcal{B}=N_T $). Indeed, if $|A|=L\le \frac{N_S}{2}$,  we shall show that 
the maximal EE $S_A=L\ln 2$   is obtained when the filling fraction $f$ is such that 
$\frac{L}{N_s}\le f \le 1-\frac{L}{N_s}$. It follows that, for fixed $N_S$, the maximal 
EE is obtained for $L=\frac{N_S}{2}$ and $f=\frac{1}{2}$. 

For simplicity we set $|A| \leq 
|\bar{A}|$,
the case $|A| \leq |\bar{A}|$ can be similarly 
worked out by exchanging $A$ and $\bar{A}$.
Since the number of nonzero eigenvalues of the hermitean
operator $\Gamma$ is smaller or equal than $|A|$
and the maximum contribution of each of these eigenvalues
to the total EE is  $\ln{2}$, we conclude
that, in order to obtain the maximum EE, 
 $\Gamma$ should
have $|A|$ eigenvectors $\vert \alpha_i \rangle$, $i=1,\ldots,|A|$ with
eigenvalue $1/2$.
Namely, one should have
\begin{equation}\label{eq:eigengonehalf}
\Gamma \vert \alpha_i \rangle= \left(1/2 \right)\vert \alpha_i \rangle.
\end{equation}

From the definition of $\Gamma$, one easily sees that, in order to have a maximum EE state, $|B|$ 
should be at least equal to $|A|$.
As a consequence, if the subspace $\mathcal{B}$ is spanned 
by the orthonormal vectors $\vert \beta_1\rangle, \vert \beta_2\rangle, \ldots
\vert \beta_{|B|}\rangle$, (so that $P_{\mathcal{B}}=\sum_{i=1}^{|B|}
|\beta_i\rangle \langle \beta_i|$), without loss of generality, one may choose the  first 
$|A|$ vectors $\vert \beta_i \rangle,\ldots,\vert \beta_{|A|} \rangle$ to have a nonzero
projection on $\mathcal{A}$. One has then that
\begin{equation}
\vert \beta_1\rangle = \gamma_1 \vert \alpha_1 \rangle  + \bar{\gamma}_1 \vert\bar{\alpha}_1 
\rangle, \vert \beta_2  \rangle= \gamma_2 \vert \alpha_2 \rangle+ \bar{\gamma}_2  \vert \bar{\alpha}_2 \rangle,
 \ldots, \vert  \beta_{|A|} \rangle = \gamma_{|A|} \vert \alpha_{|A|} \rangle+ \bar{\gamma}_{|A|} 
\vert \bar{\alpha}_{|A|}\rangle. \label{trial_basis}
\end{equation}
where the complex coefficients $\gamma_i,\bar{\gamma}_i$ are yet to be determined, the $\vert 
\alpha_{1} \rangle, \vert \alpha_{2} \rangle, \ldots,
\vert \alpha_{|A|} \rangle$
are an orthonormal basis for $\mathcal{A}$
and the $\vert \bar{\alpha}_1\rangle, \vert \bar{\alpha}_2\rangle, \ldots,
\vert \bar{\alpha}_{|A|}\rangle$
are orthonormal vectors in $\bar{\mathcal{A}}$.
Of course, additional vectors
will not give rise to nonzero
eigenvalues of $\Gamma$ thus we can limit ourselves to
$|B| = |A|$ i.e. we are
determining $\mathcal{B}$ up to vectors orthogonal
to $\mathcal{A}$ \cite{note}. The above decomposition indeed
is similar to the one obtained in \cite{klich06} where 
it was used in order to obtain \emph{lower
bounds} for the entanglement entropy in a fermionic 
systems.

\begin{figure}[t]
\centerline{
{\includegraphics[width=1\columnwidth]{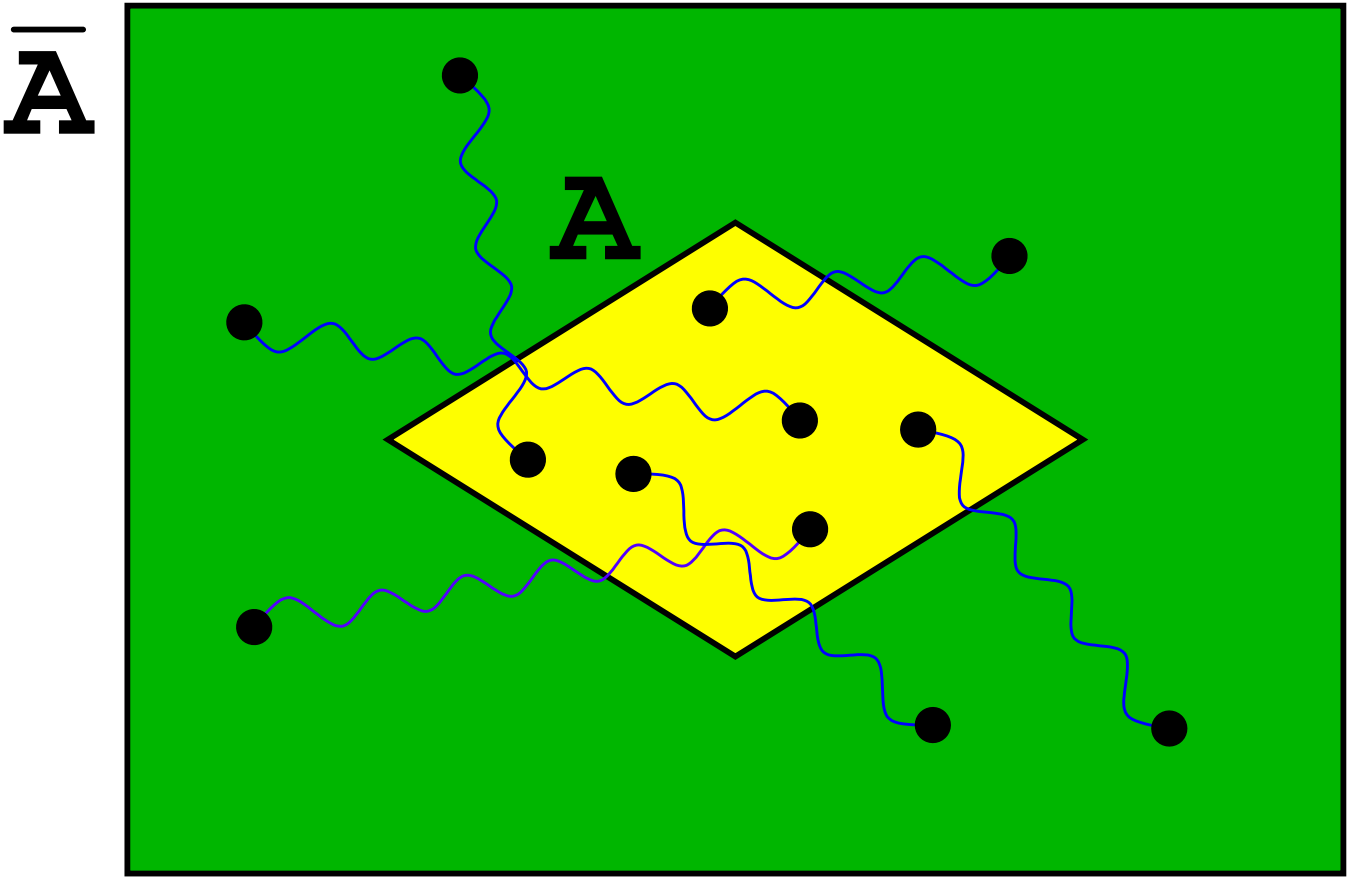}}
}
\caption{Pictorial representation of a BP-state.}
\label{fig3}
\end{figure}

By requiring that (\ref{eq:eigengonehalf}) is satisfied, one gets that
$|\gamma_i|^2=|\bar{\gamma}_i|^2=1/2$ ($i=1,\cdots, |A|$). Without loss of generality
one may choose $\gamma_i=\bar{\gamma}_i=1/\sqrt{2}$, thus $\mathcal{B}$
is spanned by:
\begin{equation}
\vert \beta_1\rangle = \frac{1}{\sqrt{2}}( \vert \alpha_1 \rangle+  \vert \bar{\alpha}_1 \rangle),
\vert \beta_2\rangle = \frac{1}{\sqrt{2}}( \vert \alpha_2 \rangle+  \vert \bar{\alpha}_2 \rangle),
 \ldots, 
 \vert \beta_{|A|}\rangle = \frac{1}{\sqrt{2}}( \vert \alpha_{|A|} \rangle
 + \vert  \bar{\alpha}_{|A|} \rangle).\label{solution_basis}
\end{equation}
This determines the form of the maximal EE state, and explicitly shows that, given $ \mathcal{A}$, 
the space $\mathcal{B}$ maximizing the EE is made out of $L$ Bell-paired states among $A$ and 
$\bar{A}$. In the following we shall refer to these states as BP-states. A BP-state is pictorially 
represented in Fig. \ref{fig3}. 

It should be stressed that 
\begin{itemize}
\item in our construction the  nature of the set $\mathcal{B}$ is left 
unspecified, the only natural 
requirement being that it is a set of allowed single particle states. 
Only for translational invariant systems the set $B$ may 
coincide with the set of single particle momentum states;
\item the set $A$ does not need to be simply connected: 
in particular, if $A$ it is not simply connected the BP-states provided by the 
indicated construction are 
not localized around the sites;
\item if, instead, the set $B$ is fixed, our construction allows to determine 
the set $A$ yielding the state with maximal EE;
\item if one wishes 
to find maximal EE states as the size of 
$A$ is enlarged, the problem to be considered is the
following: given a sequence of sets
$\{A_i\}, i=1,\ldots, N_S$ (with $A_i \subset A_j$ if $i<j$) 
to which 
corresponds a set of linear spaces $\{\mathcal{A}_i\}$
one should determine a subspace $\mathcal{B}$
for which the EE is maximal for every $i$. For this purpose it is enough 
to construct 
the basis \eqref{solution_basis} when $|A|_i = N_S/2$ for $N_S$
even or $|A|_i = (N_S-1)/2$ for $N_S$
odd: in other words the maximal EE states are obtained at half-filling and 
have $S=L \ln{2}$ for $L \le N_S/2$ as plotted in Fig.\ref{fig2}, red line;
\item the BP-state also maximizes the  R\'enyi
entropy of order $\nu$ since, for $L \le N_S$,
\begin{equation}
S_\nu = \frac{1}{1-\nu} \ln \sum_{\gamma=1}^{L}  \left[ \left(C_\gamma\right)^\nu 
 + \left(1-C_\gamma\right)^\nu \right].
\end{equation}
As one can easily verify that the R\'enyi
entropy of order $\nu$ is bounded from above by $S_\text{max}$ and 
attains a maximum when all the $C_\gamma$
are equal to $1/2$: therefore, for a BP-state,  all the 
R\'enyi entropies are maximal and equal to the maximum value of the 
EE $S_{max}=L \ln{2}$. 
\end{itemize}

\begin{figure}[t]
\centerline{
{\includegraphics[width=1\columnwidth]{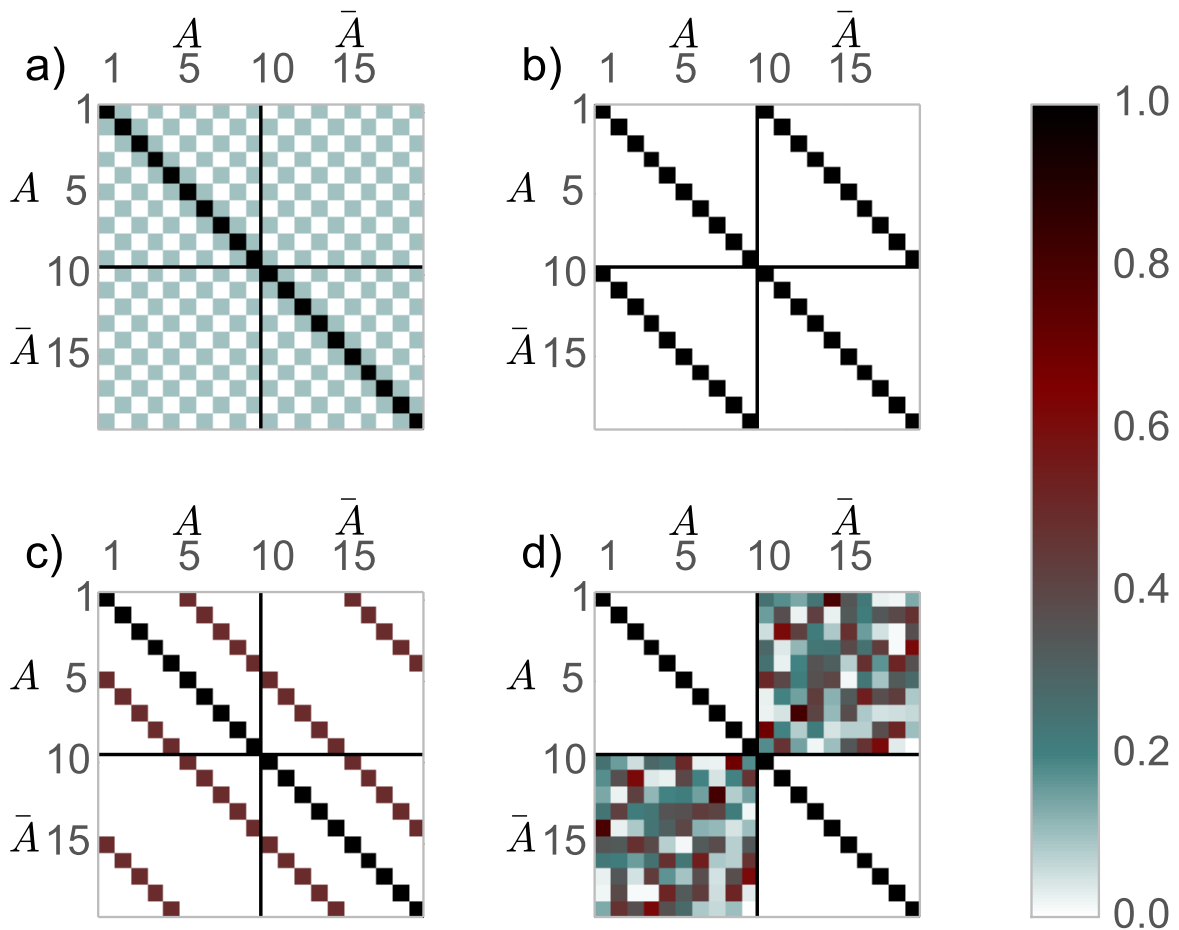}}
}
\caption{Absolute value of the matrix elements of the correlator
$\mathcal{C}_{I,J}$ for four different kind of entangled states
at half-filling in a system with $N_S=20$. The four matrices
$a$, $b$, $c$, and $d$ refer to a state with ferromagnetic
hopping, zigzag state, a state with two filled states and two holes
alternatively and a generic BP-state respectively.}
\label{fig4}
\end{figure}

Let us consider now a spatial partition in which $A$ is a simply connected 
subsystem of the lattice:  
then the states in $A$ sharing
the Bell-pairs with $\bar{A}$ are localized 
around sites. One then expects that,  plotting $C_{i,j}$ 
($i,j \in A$) for a maximal 
EE state, yield   $C_{i,j}=1/2 \delta_{i,j}$. A useful quantity to visualize 
the correlations emerging between $\mathcal{A}$ and $\bar{\mathcal{A}}$ is the correlator $\mathcal{C}_{I,J}=
\langle c^\dagger_I c_J + c^\dagger_J c_I\rangle$ which
equals $C_{I,J}+C_{J,I}^\ast$ if $I$ and $J$ belong to $A$.
We plot $|\mathcal{C}_{I,J}|$ for various
states in Fig.\ref{fig4}, where 
the correlation matrix $\mathcal{C}_{i,j}$ of $A$ 
is the one on the top left part of $\mathcal{C}_{I,J}$.

The translationally invariant states obtained 
by occupying contiguous momentum eigenvectors up to the Fermi wave vector 
are characterized by an alternating pattern
of zero and nonzero diagonals as we move away
from the main diagonal (which is by construction
equal to one), as shown in panel (a) of Fig.\ref{fig4}.
The maximal EE state for translational
invariant states is plotted in panel (b) and 
it is made out of four identity submatrices.
It should be noticed that the BP- state is made out 
of Bell pairs connecting sites at distance $N_S/2$.

The panel (c) instead refers to a state where
the occupation in $k$-space alternates in the momentum space sequences of 
two filled states and two holes 
(in Fig.\ref{fig2} its EE is represented
with the continuous scarlet line and it does not have maximal EE). 
Finally panel 
(d) shows the general structure
of a BP-state having maximal EE obtained coupling
$N_S/2$ randomly chosen orthogonal states
$\{\vert \alpha_i \rangle\}$ in subspace $\mathcal{A}$ and
$N_S/2$ randomly chosen orthogonal states
$\{\bar{\alpha}_i\}$ in subspace $\mathcal{\bar{A}}$.

Notice that, by a change of basis, the maximal 
EE state plotted in  panel (d) may be represented as in panel (b). One 
concludes that the condition that the  top left matrix of the matrix 
$\mathcal{C}_{I,J}$ is diagonal is sufficient to have a maximal EE state.

\section{Models violating the area law: Explicit Hamiltonians and their Fermi surfaces}\label{hamiltonians}

We shall exhibit here a few one-dimensional models and supporting ground-states 
leading to a violation of the area law, and we shall look 
at the nontrivial topology of the Fermi surface.

For this purpose we  firstly notice that, for a translationally invariant chain 
and for $A$ a simply connected domain such that $\vert A \vert=N_S/2$, the maximum EE state may be obtained 
by occupying the even (or odd) momentum eigenvectors. Indeed, a basis   of  $\mathcal{A}$ 
is given by vectors $\vert \alpha_k \rangle$  such that
\begin{equation}\label{alpha}
\langle J | \alpha_k \rangle = \left\{ 
 \begin{array}{ll} 
  \frac{1}{\sqrt{N_S/2}} e^{ i k J} & \text{for }J\leq N_S/2 \\
  0 & \text{for }J > N_S/2,
 \end{array}\right.
\end{equation}
while a basis of  $\bar{\mathcal{A}}$ is given by vectors  $\vert \bar{\alpha}_k \rangle $  such that
\begin{equation}\label{alpha1}
 \langle J | \bar{\alpha}_k \rangle = \left\{ 
 \begin{array}{ll}
  0 &  \text{for }J\leq N_S/2 \\
  \pm \frac{1}{\sqrt{N_S/2}} e^{ i k J} & \text{for }J > N_S/2.
 \end{array}\right.
\end{equation}
In Eqs. (\ref{alpha1}) and (\ref{alpha})  $k=4 \pi n_k/N_S$ with $n_k=-N_S/4,\cdots,N_S/4-1$ and
the $\pm$ accounts for the filling of the
even and odd frequencies respectively.
One sees that the subspace spanned
by \eqref{solution_basis} is the state with 
only the even (odd) frequencies occupied. This is a BP-state maximizing the EE.  

Notice that, if we choose a state 
with an alternating sequence of two filled momenta and two holes, then 
the EE is linear up to $L=N_S/4$: for $N_S=12$ this corresponds to 
the EE thick scarlet line of Fig.\eqref{fig4}, i.e., the 
$15$th state from top in the right 
part of the legend of the same figure. Similarly, for the state 
with an alternating sequence of three filled momenta and three holes, 
the EE is linear up to $L=N_S/6$: for $N_S=12$ this corresponds to 
the EE thick green line of Fig.\eqref{fig4}, i.e., the bottom state in the left 
part of the legend. 
For general $N_S$ at half-filling, a state having a  
sequence of $n$ filled momenta and $n$ holes 
(with $N_S$ multiple of $2n$) will have 
linear EE up to $L=N_S/2n$.

A simple Hamiltonian  supporting a BP-state as a ground state, thus yielding the volume law for  the EE, has 
the  form \eqref{Hamiltonian}
with a hopping matrix $t_{I,J}$ given by (with even $N_S$) 
\begin{equation}
t_{I,J} = \left\{ 
 \begin{array}{ll}
  -t & \quad for \, |I-J|_{p}=\frac{N_S}{2} \\
  0 & \quad otherwise,
 \end{array}\right.
\label{example}
\end{equation}
with $t>0$ and periodic boundary conditions. Notice that, in \eqref{example}, 
only hoppings between sites distant $N_S/2$ are allowed. At half-filling,  
the ground-state is constructed by occupying 
only the states with $n_k$ even (occupation of the states with $n_k$ odd 
is obtained for $t<0$). As a result, the  Fermi  surface  has a fractal 
topology and its counting  box dimension $d_{box}$ \cite{falconer13} 
equals to $1$. This example provides  an 
explicit  and simple setting where the emergence of the volume law 
is associated to a non-trivial topology of the Fermi 
surface: this sheds light on the results of previous investigations
\cite{wolf06,huijse13}.
Slight modifications of the hoppings \eqref{example} 
can be built to have states having 
sequences of $n$ filled momenta and $n$ holes.

Fractal  Fermi surfaces may be realized as pertinent  limits 
of other model Hamiltonians. In the following  we analyze two specific 
models where this happens.

\subsection{Model A}\label{long_phi}
A possible way to obtain a fractal Fermi surface is 
to consider the effect of a phase  
in a model with long-range hoppings: 
\begin{equation}
t_{I,J}=\frac{t \cdot e^{i \phi d_{I,J}  }}{|I-J|_{p}^\alpha},
\label{magnetic_phase}
\end{equation}
where $\phi=\frac{2\pi}{N_S} \Phi$, being $\Phi$ a constant and 
$d_{I,J}$ the oriented distance between the sites $I$ and $J$, 
whose definition is given in (\ref{eqn:distphase}).

The spectrum of the ensuing hopping Hamiltonian is analyzed in 
Appendix \ref{app:LRphase}. For odd $N_S$ the eigenvalues are given by 
$\varepsilon_k=-2t\ell_\alpha(k;N_S)$ 
where
\begin{equation}
\ell_{\alpha}(k;N_S)=\sum_{m=1}^{(N_S+1)/2}\frac{\cos\left[ m \left( k +\phi 
\right)\right]}{m^\alpha};
\label{elle}
\end{equation}
as usual, $k=2\pi n_k/N_S$ with $n_k=0,\cdots,N_S-1$. A similar formula 
is obtained for even $N_S$.

\begin{figure}[t]
\centerline{
{\includegraphics[width=1\columnwidth]{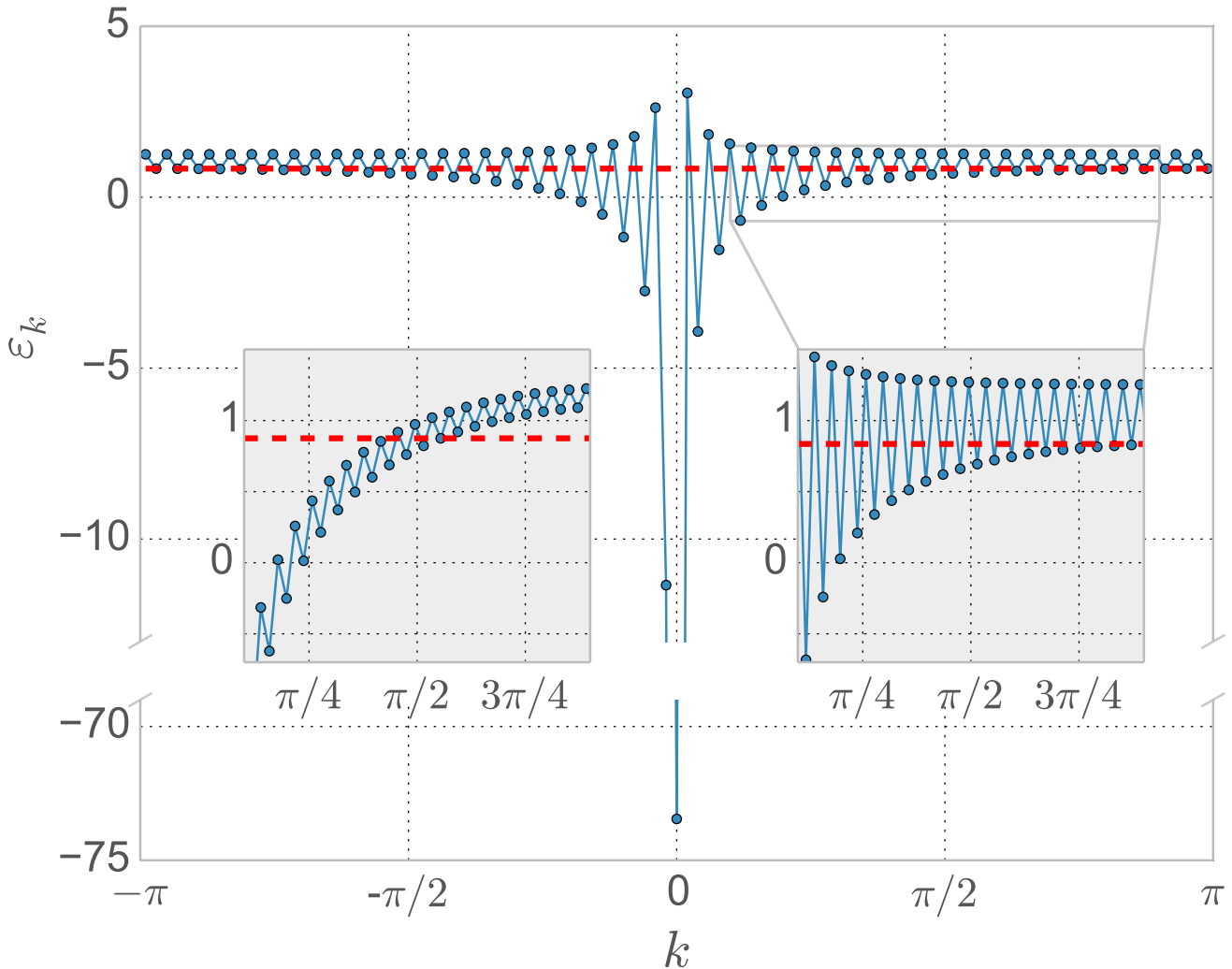}}
}
\caption{Spectrum of the long-range Hamiltonian \eqref{magnetic_phase}   
with $\Phi=0.1$, $\alpha= 0.1$, filling factor $f=0.5$ and $N_S=100$.
Right inset: detail of the main plot showing the
alternating occupation of the modes $k$, the Fermi energy corresponding 
to the dashed line.
Left inset: loss of the alternating occupation increasing $\alpha$, with 
$\Phi=0.1$, $\alpha=0.4$, $f=0.5$ and $N_S=100$.}
\label{fig5}
\end{figure}

\begin{figure}[t]
\centerline{
{\includegraphics[width=1\columnwidth]{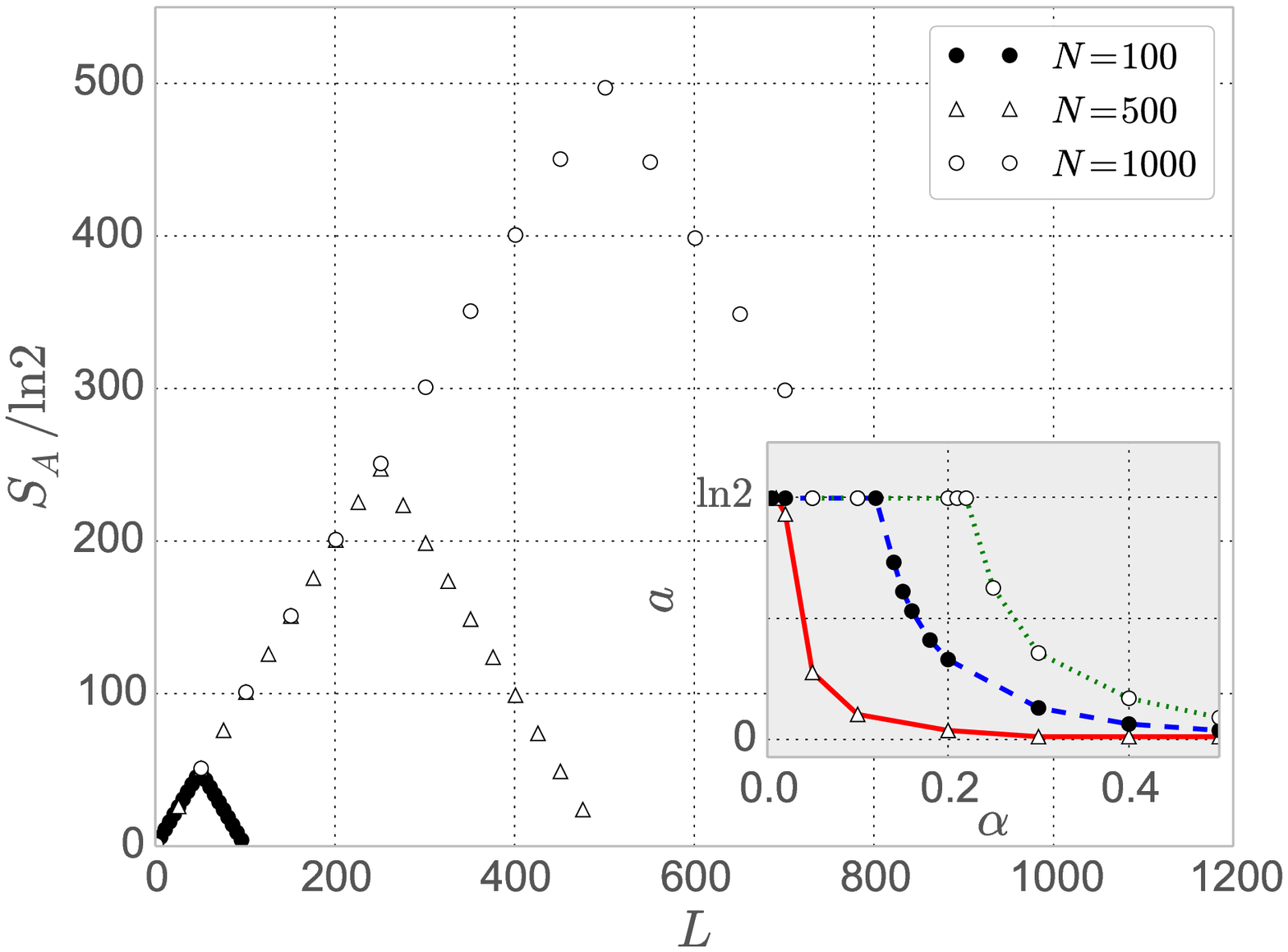}}
}
\caption{Entanglement entropy as a function of the size 
of the block with $\alpha=0.1$, $\Phi=0.1$. Different total number of sites are considered: 
$N_S=100$ (full circles), $N_S=500$ (triangles), $N_S=1000$ (empty circles).
Inset: Slope of the entanglement entropy fitted with a linear function $S=a L+b$ for different values of $\alpha$:  
$\Phi=0.01$ (triangles), $\Phi=0.1$ (full circles), $\Phi=0.3$ (empty circles) and $N_S=300$.
}
\label{fig6}
\end{figure}

For $\phi=0$, the spectrum is always monotonous in the interval 
$k \in [0 ,\pi ] $ , while for $\phi>0$ the spectrum 
is monotonous for $\alpha \ge 1$. More precisely, at fixed 
$\phi$ and $N_S \gg 1$, there is a critical 
value of $\alpha_c$, depending both on $N$ and $\phi$, 
such that, for $\alpha<\alpha_c$, at half-filling, 
all the momenta $k$ 
are  occupied in an alternating way, as shown in Fig.\ref{fig5}. Thus,  
for $\alpha<\alpha_c$ and at half-filling, 
 the ground-state is 
a BP-state, EE is linear with slope $\ln{2}$ and the Fermi surface has a fractal 
 topology with $d_{box}=1$: this is shown in 
Fig.\ref{fig6}. 

For $N_S \to \infty$ one has that  $\alpha_c \to 0$: this 
happens since, in the thermodynamic limit ($N_S=\infty$), 
it is not possible to define and occupy only the even momenta; however,  
for each $N_S$ arbitrarily large, $\alpha_c$ is strictly positive. When 
$\alpha>\alpha_c$ only a fraction of the momenta are occupied in an 
alternating way, since the ``zig-zag'' structure of the dispersion relation is partially lost. As a result, 
the slope of the EE decreases, as shown in the inset 
of Fig.\ref{fig6}.

\subsection{Model B}\label{accumul}
\begin{figure}[t]
\centerline{
{\includegraphics[width=1\columnwidth]{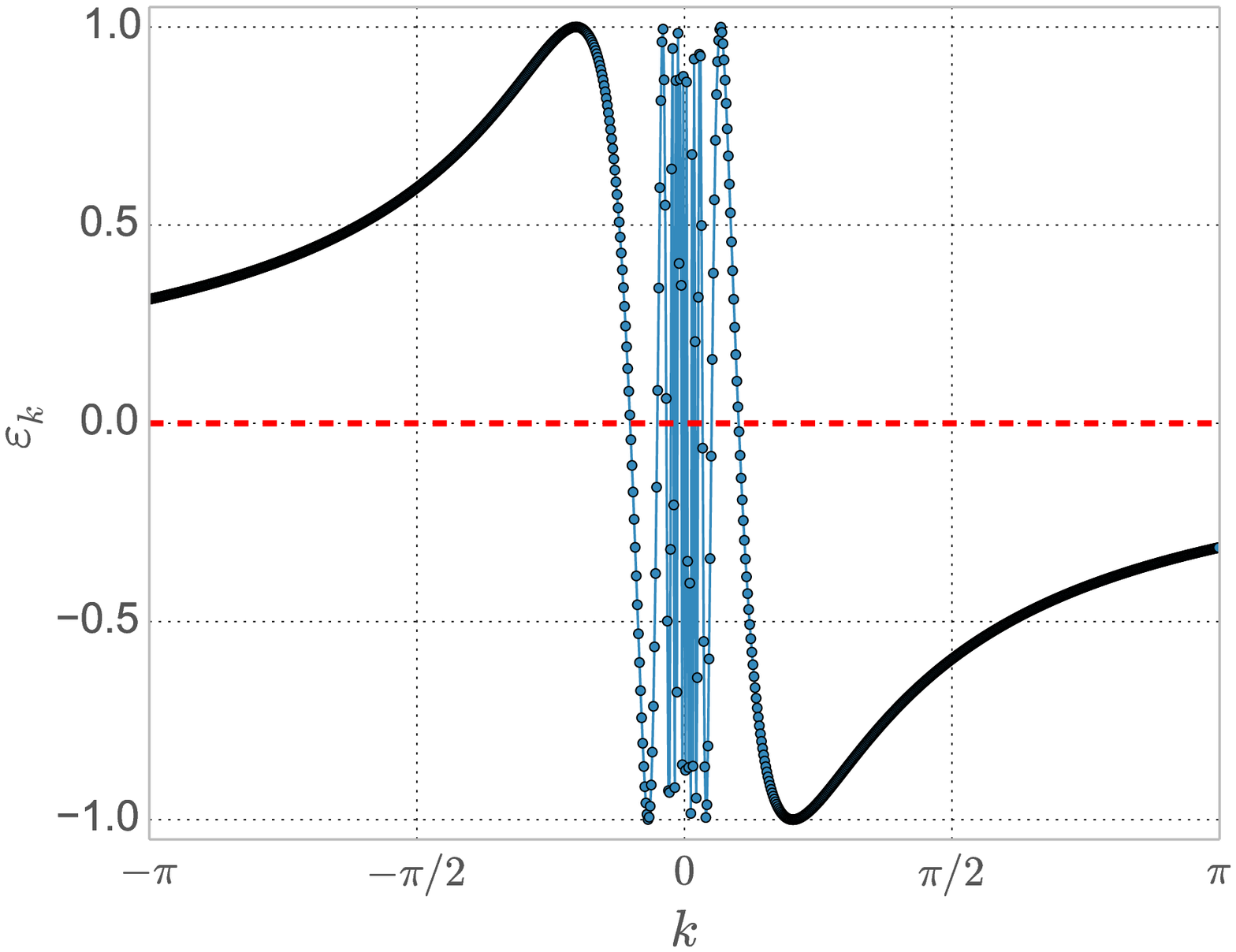}}
}
\caption{Energy spectrum corresponding to \eqref{sin_1_k} and $\alpha=1$, 
with the thick dashed line corresponding to the Fermi energy for half-filling.}
\label{fig7}
\end{figure}

We consider here a translational invariant 
chain of $N_S$ sites with periodic boundary conditions and 
with eigenfunctions given by plane waves 
$\psi_k(J)=\frac{1}{\sqrt{N_S}} e^{ikJ}$. We assume that 
the Hamiltonian is such that the single-particle energy spectrum has the form 
\begin{equation}
\varepsilon_k=-t \cdot \sin{\left( \frac{1}{k^\alpha} \right)},
\label{sin_1_k}
\end{equation}
where $\alpha$ is a positive odd integer. 
The spectrum is plotted in Fig.\ref{fig7} for $\alpha=1$. The Fermi surface, in a pertinent 
range of fillings, has a fractal  topology and, at half-filling, the Fermi energy 
is zero so that Fermi surface is simply given by the set of  points 
$\{\pm\frac{1}{\pi \alpha}, \pm\frac{1}{\pi 2 ^\alpha}
\pm\frac{1}{\pi 3^\alpha} \ldots \}$.
The point $k=0$ is an accumulation point for this set
with  box counting dimension 
\cite{falconer13} 
\begin{equation}
d_{box}=\frac{\alpha}{\alpha+1},
\label{bcd_sin}
\end{equation}
so that $d_{box}=1/2$ for $\alpha=1$. 

We numerically determined the EE for the above model \eqref{bcd_sin} 
for increasing values of $N_S$ as the parameter $\alpha$
takes the values $\alpha= 1, 3, 5$.
The EE has a well defined thermodynamic limit
and, for small size of the subsystem $A$, is well described
by a power-law. In order to give a reliable estimate
of this power-law growth one needs to compute the EE
for $L=1,\ldots,128$ and fit the obtained values with the function
$S_A=a + b L^\beta$ for different system sizes.  One needs this procedure
to get rid of finite-size effects since, even in the
the short-range model, the EE shows 
finite-size effects when $L$ is comparable
to the system size  \cite{jphysa09}. 
Thus, to recover the expected logarithmic growth 
one has to fix $L$ and vary $N_S$.  

\begin{figure}[t]
\centerline{
\scalebox{0.45}{\includegraphics{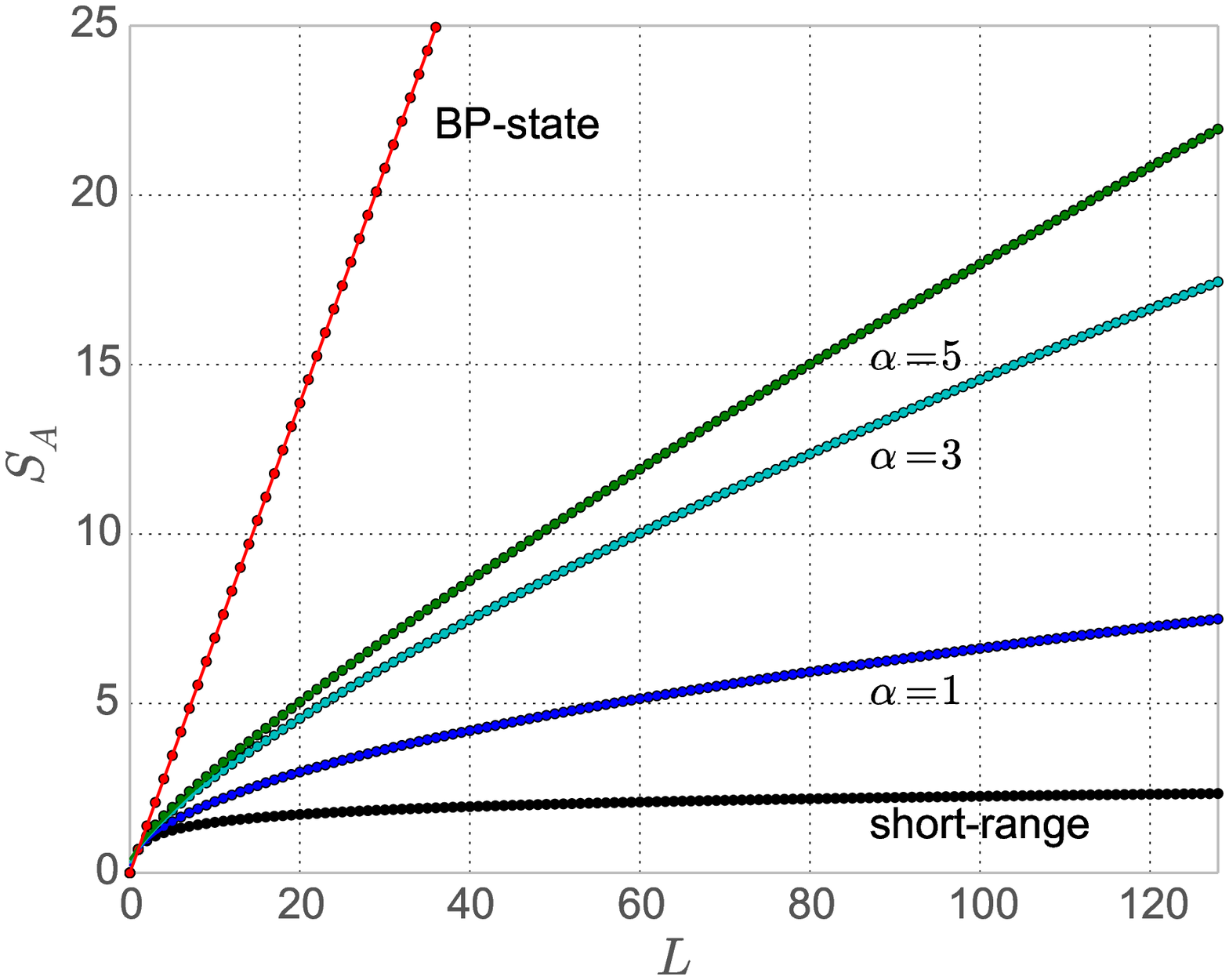}}
\scalebox{0.45}{\includegraphics{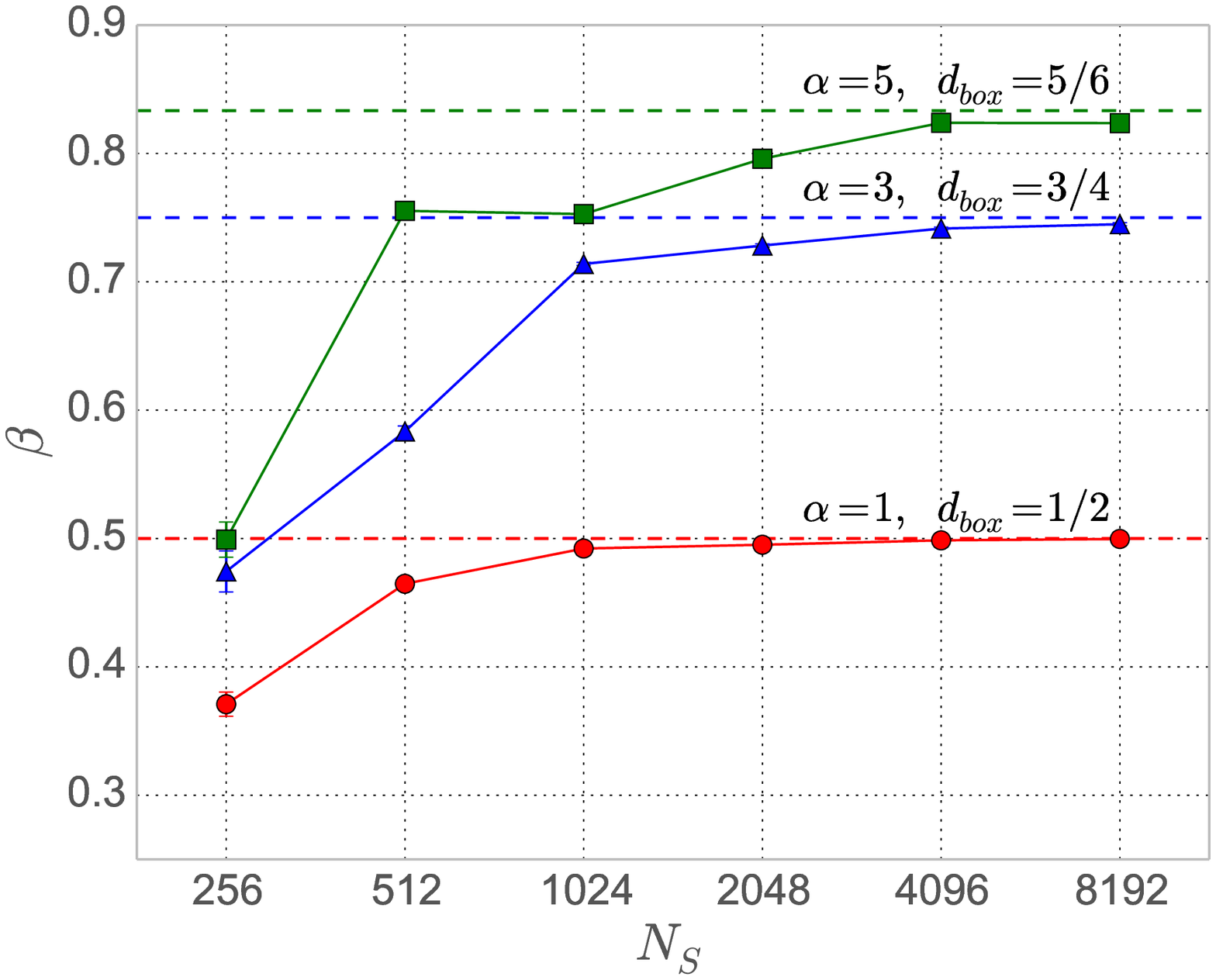}}
}
\caption{Left panel:  EE
for the model \eqref{sin_1_k} for $\alpha=1,3,5$
and $N_S=8192$.  The continuous
lines are the best fit lines (hardly distinguishable
from the numerical points).
The curves are compared with the he short-range
case, exhibiting an area law, and linear behavior of a
 ``zigzag'' filling.   
In the short-range case the continuous
curve is the best fit $S(L)=a_I + b_I \ln(L)$ 
where $a_I=0.71971$ and $b_I=0.334859$;
notice how the coefficient in front
of the logarithm is very close to the
expected value of $1/3$.
Right panel:  best fit
for the coefficient $\beta$ (with 
$S(L)=a + b L^\beta$) for various system sizes and for the three values
of $\alpha=1,3,5$  at half-filling (circles,
triangles and squares respectively). The straight lines
are given by the box counting fractional
dimension of the Fermi surface of the model.}
\label{fig8}
\end{figure}

The results of this fit are reported  in Fig. \ref{fig8}. 
Here we plot, on the left panel,  the EE for
different values of $\alpha$. The results for the EE obtained for the
 short-range model and the BP-state 
are also shown for comparison. 
In the right panel of Fig. \ref{fig8} we plot the 
exponent $\beta$ as a function of $N_S$.
We see that, as $N_S$ is increased, $\beta$ approaches $d_{box}$. 
Since this feature is shared also by the previous models, one is 
 tempted to conjecture that this is  a general feature 
of the scaling of the EE in translational invariant chains. 

Our findings agree with the 
results presented in the Appendix of \cite{huijse13}, where a construction 
of a translation invariant fermionic state 
violating the area law 
was explicitly given with Cantor-like Fermi surfaces, 
and with the results of \cite{fannes03} where, for infinitely many intervals 
in a spin chain, an EE of the form $S \sim L^\alpha$ was found with 
$\alpha$ possibly 
taking any value between $0$ and $1$. 

We observe that 
in \cite{gioev06} a formal criterion for the growth of the EE in presence 
of fractal Fermi surfaces has been discussed: in particular, it was shown in
that if $C_1 \|h\|^{\beta_\Gamma}<{\rm Vol}( \Gamma\setminus(\Gamma+h))<C_2
\|h\|^{\beta_\Gamma}$ for a small set $\|h\|$ and 
$0<\beta_\Gamma\leq1$ (with $C_1$ and $C_2$ real constants and $\Gamma$ the 
Fermi surface), 
then there is a deviation from the area law with exponent 
$1-\beta_\Gamma$ (see also \cite{fannes03,gioev02}). 
Our results imply that such coefficient $\beta_\Gamma$ for this class 
of Hamiltonians is related to the box counting dimensions through 
$1-\beta_\Gamma=d_{box}$: an interesting problem for future research 
is the study of such relation in the general case.

So far we analyzed only models in which a Fermi surface can be defined: it  is natural 
to expect that violations of the area law may emerge also in situations 
where it is impossible to define a Fermi surface. When disorder is present such situation arises  
naturally in single realizations of disorder. In Appendix \ref{sec:random} we shall analyze 
a model with random long-range hoppings and we shall see how deviations 
from the area law may appear also in absence of the translational invariance.

\section{Concluding remarks}\label{concl}
We investigated how the area law for the entanglement entropy (EE)  
may be violated in non-interacting fermionic lattices and provided a  method enabling to construct 
the states with maximal EE exhibiting a volume law. We called these states BP-states. 
For these states the EE is linear in the size of the subsystem $A$ and the Fermi surface 
has fractal topology. 

For translational invariant free fermionic Hamiltonians, BP-states may be obtained, at half 
filling, by occupying, according to Fermi statistics, even or odd momentum eigenvectors  providing 
an explicit momentum representation of the state with maximal entanglement entropy. By 
this procedure, one originates a ``zig-zag'' structure of the dispersion relation leading  for fermionic chains to 
the emergency of a fractal Fermi surface with box-counting dimension $1$. By means of this 
procedure one can construct an explicit Hamiltonian whose ground-state supports exactly 
the volume law.

We then provided some examples 
of fermionic models for which the ground state may have an EE $S_A$ between 
the area and the volume law, and we gave an explicit example of a one-dimensional free fermion model 
where the EE is such that $S_A(L)=a+b L^\beta$ with $\beta$ being intermediate between 
$\beta=0$ (area law) and $\beta=1$ (BP-state). We saw that, also for this model, the 
dispersion relation has a ``zig-zag'' structure leading to a fractal Fermi surface whose 
counting box dimension equals, for large lattices, $\beta$. 

It is attractive to speculate that there may be a general relation between the fractal 
dimension of the Fermi surface, measured by the box counting dimension, and the exponent 
$\beta$ measuring  the amount of violation of the area law for one-dimensional translational
invariant free fermion lattices.  Here, we only report the fact that, in all the one-dimensional
examples analyzed in this paper, this relation holds true. 

As a by product, our analysis shows that, a volume law for the EE cannot emerge in free 
fermion lattices as a result of long range hopping alone. Indeed, our analysis shows that, 
at least for translational invariant systems, a fractal structure of the Fermi surface is 
needed to establish a volume law for the EE.

Although we studied only non-interacting fermions on the  lattice, 
our analysis is relevant also for spin models admitting a fermionic representation.
Indeed, it has been recently 
exhibited a spin chain model supporting  a volume law 
for the EE \cite{lundgren14}: in its fermionic 
representation, the Hamiltonian is highly non-local in agreement with
the scenarios presented in this paper.

{\em Acknowledgements:} We would like to thank A. Bayat, F. Buccheri, J. Magan and E. Tonni 
for useful discussions. During the completion of this work, 
we became aware of results on volumetric law in fermionic lattices 
with inhomogeneous nearest-neighbor hoppings 
by G. Ram\'irez, J. Rodr\'iguez-Laguna, and G. Sierra \cite{ramirez14-b}: 
it is a pleasure to thank them for discussions and useful correspondence.
P.S. thanks the
Ministry of Science, Technology and Innovation of Brazil
for financial support and CNPq for granting a ``Bolsa de
Produtividade em Pesquisa''.
P.S. and S.P. acknowledge
partial support from MCTI and UFRN/MEC (Brazil).
A.S. acknowledges support from CNPq, and from The Center for Nanoscience and Nanotechnology at Tel Aviv University and the PBC Indo-Israeli Fellowship. 
A.T. acknowledges support from the Italian PRIN
``Fenomeni quantistici collettivi: dai sistemi fortemente correlati ai simulatori quantistici''
(PRIN 2010\_2010LLKJBX).
S.P. is supported by a Rita Levi-Montalcini fellowship of MIUR.
A.T. acknowledge hospitality from 
International Institute of Physics (Natal) 
where part of this work has been carried out. G.G. and A.T. acknowledge 
support from the European Project MatterWave.

{\em Note added:} After the submission of this paper we noticed on the \verb|arXiv| a very 
interesting paper on the power law violation of the area law in quantum spin chains \cite{movassagh14}.

\appendix

\section{Correlation functions and entanglement entropy}\label{app:A}
Here we derive the correlation matrix and the EE of the subsystem $A$ for a generic  graph  
${\cal G}$ with $N_S$ sites and $N_T$  fermions 
hopping on it. The model is described by the 
Hamiltonian \eqref{Hamiltonian} and the filling is $f=N_T/N_S$ with $0\le f \le 1$.
 
As in  Sec. \ref{sec1}, the sites of the lattice 
${\cal G}$ are denoted by capital letters and the
eigenvalues' equations of \eqref{Hamiltonian} read as  
\begin{equation}
-\sum_{J=1}^{N_S}  t_{IJ} \psi_\Gamma(J)=\epsilon_\Gamma \psi_\Gamma(I),
\label{eigen}
\end{equation}
where  $\epsilon_\Gamma$ are $N_S$  single-particle energy eigenvalues  ordered so that
$\epsilon_1 \le \epsilon_2 \le \cdots \le \epsilon_{N_S}$ and $ \psi_\Gamma(I)$ are the 
corresponding $N_S$ orthonormal eigenfunctions. 
Upon defining 
\begin{equation}
d_\Gamma=\sum_{I=1}^{N_S}  \psi_\Gamma(I) c_I,
\label{d}
\end{equation}
one immediately sees that the operators $d_\Gamma$ obey to the canonical fermionic anticommutation 
relations and that the Hamiltonian \eqref{Hamiltonian} may be rewritten as\begin{equation}
H=\sum_{\Gamma=1}^{N_S}
\epsilon_\Gamma d_\Gamma^{\dag} d_\Gamma,
\label{ham_d}
\end{equation}
so that the  ground-state
$\vert \Psi \rangle$ can be written as
\begin{equation}
\vert \Psi \rangle=\prod_{\Gamma=1}^{N_T} d_{\Gamma}^\dag \vert 0 \rangle.
\label{GS}
\end{equation}

Given a set $A$, whose sites are labelled by  $i,j=1,\cdots,L$, one may define the correlation 
matrix $C$ as the matrix whose entries  are given by
\begin{equation}
C_{ij}=\langle \Psi \vert c_i^\dag c_j \vert \Psi \rangle.
\label{C}
\end{equation}
Using \eqref{d} and \eqref{GS}, one finds
\begin{equation}
C_{ij}=\sum_{\Gamma=1}^{N_T} \psi_{\Gamma}(i) \psi_{\Gamma}^\ast (j).
\label{C-eigen}
\end{equation}
If one denotes with  $C_\gamma$ ($\gamma=1,\cdots,L$)  the $L$ eigenvalues of the matrix 
$C_{ij}$ one gets \cite{peschel09}  
\begin{equation}
S_{A}=-\sum_{\gamma=1}^{L} \left[ \left( 1-C_\gamma \right)
\ln{\left( 1-C_\gamma \right)} + C_\gamma \ln{C_\gamma} \right].
\label{res_app}
\end{equation}

\section{Spectrum of model A}\label{app:LRphase}
We analyze here the spectrum of the model, introduced in Sec.  \ref{long_phi}. The hopping matrix reads
\begin{equation}
t_{I,J}=t\frac{e^{i\phi d_{I,J}}}{\left|I-J\right|_{p}^{\alpha}},
\end{equation}
where
\begin{equation}\label{eqn:distphase}
d_{I,J}=\begin{cases}
(I-J) & if\:\left|I-J\right|\leq N_{S}-\left|I-J\right|\\
-N_{S}+\left|I-J\right| & otherwise.
\end{cases} 
\end{equation}
Due to the translational invariance, the eigenstates are 
 plane waves, while, for finite $N_{S}$,
the spectrum is given by
\begin{equation}
\varepsilon_{k}=-2t\begin{cases}
\mbox{\ensuremath{\sum_{m=1}^{\frac{N-1}{2}}\frac{1}{m^{\alpha}}\cos\left((k+\phi)m\right)}} & for\: odd\: N_{S}\\
\sum_{m=1}^{\frac{N}{2}-1}\frac{1}{j^{\alpha}}\cos\left(\left(k+\phi\right)m\right)+
\frac{\cos\left(\pi n_{k}\right)}{2\left(\frac{N_S}{2}\right)^{\alpha}} & for\: even\: N_{S},
\end{cases}
\label{lerch}
\end{equation}
with $k=2\pi  n_k /N_S$. Even if, for finite $N_S$, $\varepsilon_k$ forms a discrete set 
corresponding to integer values of $n_k$, it is most useful to provide an expression of 
(\ref{lerch}) valid for all values of $k$. For this purpose,   
Eq. \eqref{lerch} may be rewritten using Lerch transcendent
functions \cite{lerchzeta}: 
\begin{equation}
\Phi(z,\alpha,a)=\sum_{j=0}^{\infty}\frac{z^{j}}{\left(j+a\right)^{\alpha}},
\end{equation}
yielding
\begin{equation}\label{eqn:eklerch}
\varepsilon(k)=2t\begin{cases}
\Re\left[z^{\frac{N+1}{2}}\Phi(z,\alpha,\frac{N+1}{2})-z\Phi(z,\alpha,1)\right] & for\: odd\: N_{S}\\
\Re\left[z^{\frac{N}{2}}\Phi(z,\alpha,\frac{N}{2})-z\Phi(z,\alpha,1)\right]
-\frac{\cos\left(\pi n_{k}\right)}{2\left(\frac{N}{2}\right)^{\alpha}} & for\: even\: N_{S},
\end{cases}
\end{equation}
with $z \equiv e^{i\left(k+\phi\right)}$. 

Let us start by analyzing the spectrum when $\phi=0$;  this corresponds
to power-law decaying hoppings.  Since $\varepsilon_{k}=-\varepsilon_{k}$ one may consider only
the interval of the Brillouin zone corresponding to $k=\left[0,\pi\right)$. 
For $\alpha \geq 1$, $\varepsilon(k)$ is a monotonically increasing function
of $k$, so that  the many-body ground-state is filled  following an ascending order of $\left|n_{k}\right|$,
just as in the short-range tight-binding
model.  Thus, for every value of $\alpha\geq1$, 
the EE is the same of the tight binding model and, thus,  follows the usual area law for the EE.
For $\alpha<1$, $\varepsilon(k)$ is an oscillating function with $\frac{N_{S}-(N_{S}\mod2)}{2}$
maxima (and minima) almost equidistant in the interval $k=\left[\pi,\pi\right)$.
In addition, the set $\varepsilon_{k}$ is still monotonous in
$k=\left[0,\pi\right)$ so that  every wave number $k$ lies between a different
pairs of local maxima an minima. As a result, eve for $\alpha<1$ the EE follows an area law.  

 If $\phi \neq 0$ the function
$\varepsilon(k)$ shifts, losing its parity. For $\alpha<1$, if one considers two consecutive 
$k$s of the discrete set $\varepsilon_{k}$ one sees that the
shift introduced by a small $\phi$ increases the
energy of one of them and decreases the energy of the other. It follows that the
set $\varepsilon_{k}$ is no more monotone and takes  a zigzag shape. As  shown in Fig.\ref{fig5}, 
the energies corresponding to $n_{k}$'s 
of different parity arrange themselves on two different branches.  

\begin{figure}[t]
\centerline{
\scalebox{0.45}{\includegraphics[angle=-0]{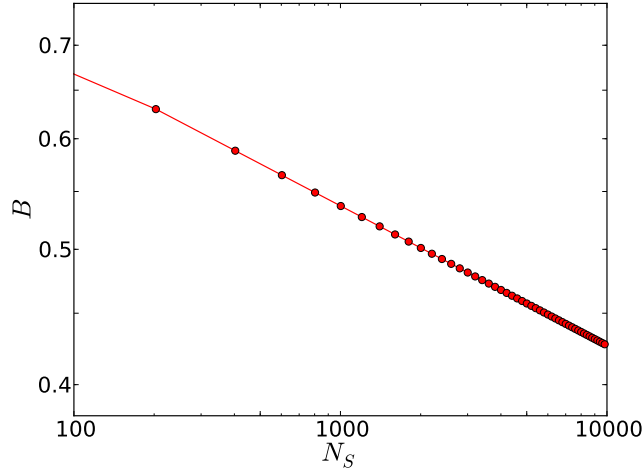}}
}
\caption{Oscillation amplitude of the spectrum \eqref{eqn:eklerch} for $k=\pi$ and $\alpha=0.1$.
}
\label{fig9}
\end{figure}

The maximum spacing between the two branches is bounded by the amplitude
of the oscillation of $\varepsilon(k)$. To give an estimate of that, one may , for  $\phi=0$, 
approximate $\varepsilon(k)$ around $k=\pi$
with a cosine function \cite{lerchzeta}
\begin{equation}
\varepsilon(k)/t\simeq A+B\cos(R(k+\delta)),
\end{equation}
where
\begin{eqnarray}
R & = & \frac{N_{S}-(N_{S}\mod2)}{2},\\
a & = & \frac{\varepsilon(\pi),}{t}\\
b & = & \frac{\varepsilon'(\pi)}{t},\\
c & = & \frac{\varepsilon''(\pi)}{2t},\\
A & = & a+\frac{2c}{R^{2}},\\
B & = & \pm\frac{2}{R^{2}}\sqrt{\frac{b^{2}R^{2}}{4}+c^{2}},\\
\delta & = & \frac{1}{R}\arctan\frac{bR}{2c}-\pi.
\end{eqnarray}
The study of the amplitude $B$ shows a weak polynomial dependence
on the number of sites, as plotted in Fig.\ref{fig9}, so that a zigzag
behavior of the spectrum is expected for every finite $N_{S}$.

The lower branch has always negative concavity in $k=\pi$, while the
top branch's concavity can be positive or negative, depending on $\alpha$.
In the first case and for half-filling, there appears an alternation in the
occupancy of $k$, i.e the Fermi energy is lying between the two branches,
giving rise to the BP-state described in section \ref{topol}. The concavity
remains positive for $\alpha<\alpha_{c}$ and, in this regime, the
EE is always maximal showing a volume law behavior with fixed slope
(see inset in Fig.\ref{fig5}). For $\alpha>\alpha_{c}$, as the concavity of the
upper branch becomes negative, some consecutive $k$s close to $k=0$
are occupied while close, to $k=\pi$ the Fermi Energy goes below
the two branches. This breaks the complete alternating configuration
but, for $\alpha$ larger but close to $\alpha_{c}$ the EE has still
depends linearly on $L$ but  with a lower slope. The crossover from volume to
area law occurs smoothly as $\alpha$ is increased.

\section{The fully connected network}\label{app:fcn}
For $\alpha=0$ the long-range hopping model \eqref{tt} becomes 
simply
\begin{equation}
H=-\frac{t}{N_S} \sum_{I\neq J}c_{I}^{\dag}c_{J},
\label{fully_conn}
\end{equation}
where we divided  the hopping coefficient $t$ by $N_S$ to keep the single particle 
spectrum lower bounded. The spectrum of (\ref{fully_conn}) is made of two eigenvalues: 
$ \epsilon_0=-t\left(N_S-1\right)/N_S$  and $\epsilon_1=t/N_S$ corresponding, 
respectively, to a non degenerate ground state and a $N_S-1$ degenerate excited state. 
The entries $\langle c_I^\dag c_J \rangle$ of the correlation matrix are given by 
\begin{equation}
\left\langle c_{I}^{\dagger}c_{J}\right\rangle =\begin{cases}
f\;\; for\; & I=J\\
b\;\; for\; & I\neq J,
\end{cases}
\end{equation}
where $f=N_T/N_S$ is the filling and $b$ has to be determined. 
Since the ground-state 
energy (more precisely, the free energy for $T \to 0$) is 

\begin{equation}
\left\langle H\right\rangle =-t\left(N_S-1\right)b=
-\frac{t}{N_S}\left(1-N_S\right)+\frac{t}{N_S}\left(N_T-1\right),
\end{equation}
it follows that
\begin{equation}
b=\frac{1-f}{N_S-1}.
\end{equation}

The correlation matrix has, the, the form 
\begin{equation}
\mathbf{C}=f\mathit{\mathbf{I}}+b\left(\begin{array}{cccc}
0 & 1 & 1\\
1 & 0 & 1\\
1 & 1 & \ddots\\
 &  &  & 0
\end{array}\right),
\end{equation}
with eigenvalues
\begin{equation}\label{eqn:eta0}
\eta_{0}=(L-1)b+f,
\end{equation}
\begin{equation}\label{eqn:eta1}
\eta_{1}=f-b.
\end{equation}
Inserting (\ref{eqn:eta0}) and (\ref{eqn:eta1}) into (\ref{res}) one gets 
\begin{equation}\label{eqn:sa}
S_{A} = -\left(1-\eta_{0}\right)\ln\left(1-\eta_{0}\right)-\eta_{0}\ln\left(\eta_{0}\right)
-(L-1)\left(1-\eta_{1}\right)\ln\left(1-\eta_{1}\right)-(L-1)\eta_{1}
\ln\left(\eta_{1}\right),
\end{equation}
which, for $N_S\rightarrow\infty$, yields
\begin{equation}\label{c9}
S_{A} \approx 
-L\left[\left(1-f\right)\ln\left(1-f\right)+f\ln\left(f\right)\right].
\end{equation}

From (\ref{c9}) a volume law for the EE is attained. However, 
$S_A$ is not a true measure of entanglement since the initial state is mixed and, in addition, 
the mutual information turns out to be zero. As a result, the emergence  of a volume law does 
not lead to nonlocal correlations for this model.

In the following, we show that the mutual information is indeed zero.
If $A$ and $\bar{A}$ are two complementary sets covering the full lattice, the mutual 
information is defined as
\begin{equation}
I\left(A:\bar{A}\right)=S\left(A\right)+S\left(\bar{A}\right)-
S\left(A \cup \bar{A}\right)=S_{A}+S_{\bar{A}}-S_{T},
\end{equation}
where $S_T$ is the total entropy. 
For a mixture of  $N_{deg}$ degenerate states, the total entropy is given by
$S_T$ 
\begin{equation}
S_{T}=-\ln{\frac{1}{N_{deg}}}.
\end{equation}
Here, the degeneracy of the many-body ground state is given by 
\begin{equation}
N_{deg}=\binom{N_S-1}{N_T-1},
\end{equation}
so that
\begin{equation}
S_{T}=-\left[\ln\left(N_S-1\right)!-\ln\left(N_T-1\right)!-\ln\left(
N_S-N_T\right)!\right].
\end{equation}
In the limit of large $N_S$, at fixed filling, one easily obtains
\begin{equation}\label{eqn:st}
S_T \approx N_S\left[\left(1-f\right)\ln\left(1-f\right)+
f\ln\left(f\right)\right].
\end{equation}
The entropy of the set $\bar{A}$ has an  analogous expression to (\ref{eqn:sa}) 
\begin{equation}\label{eqn:sbara}
S_{\bar{A}} = -\left(1-\bar{\eta}_{0}\right)\ln\left(1-\bar{\eta}_{0}\right)-
\bar{\eta}_{0}\ln\left(\bar{\eta}_{0}\right)-
(N_S-L-1)\left(1-\bar{\eta}_{1}\right)\ln\left(1-\bar{\eta}_{1}\right)-
(N_S-L-1)\bar{\eta}_{1}\ln\left(\bar{\eta}_{1}\right),
\end{equation}
where $\bar{\eta}_0=(N_S-L-1)\frac{1-f}{N_S-1}+f$ and 
$\bar{\eta}_1=f-\frac{1-f}{N-1}$. For  large $N_S$ (\ref{eqn:sbara}) becomes
\begin{equation}
S_{\bar{A}}\approx-\left(N_S-L\right)\left[\left(1-f\right)\ln\left(1-f\right)+f\ln\left(f\right)\right].
\end{equation}
Finally, putting together (\ref{eqn:sa}), (\ref{eqn:sbara}) and (\ref{eqn:st}) one gets
\begin{equation}
I\simeq0.
\end{equation}

\section{Random hopping model}\label{sec:random}
Here, we present  some
preliminary results on a model where long-range randomness is included in the hopping matrix. 
EE has been studied for different disordered models \cite{refael04,mondragon13,ramirez14,pouranvari14}; 
in particular, a violation of the area law has been found for free fermionic models in 
their metallic phase  \cite{pouranvari14}.

The model  is defined by  Hamiltonian \eqref{Hamiltonian} 
with a random long-range hopping matrix given by
\begin{equation}
t_{I,J}=\frac{t \cdot \eta_{I,J}}{|I-J|_{p}^\alpha};
\label{random}
\end{equation}
in Eq. (\ref{random}) $\eta_{I,J}$ is a random variable assuming the values 
$\pm 1$ with equal probability. 
This model breaks the translational symmetry; as a result, one cannot analyze the states 
using momentum eigenvectors. A direct diagonalization of the 
matrix $t_{I,J}$ is needed to compute the correlation 
matrix \eqref{correl_intr}, whose eigenvalues are then used 
to compute the EE of the ground state for various  sizes, $L$, of the subsystem $A$.

\begin{figure}[t]
\centerline{
\scalebox{0.45}{\includegraphics{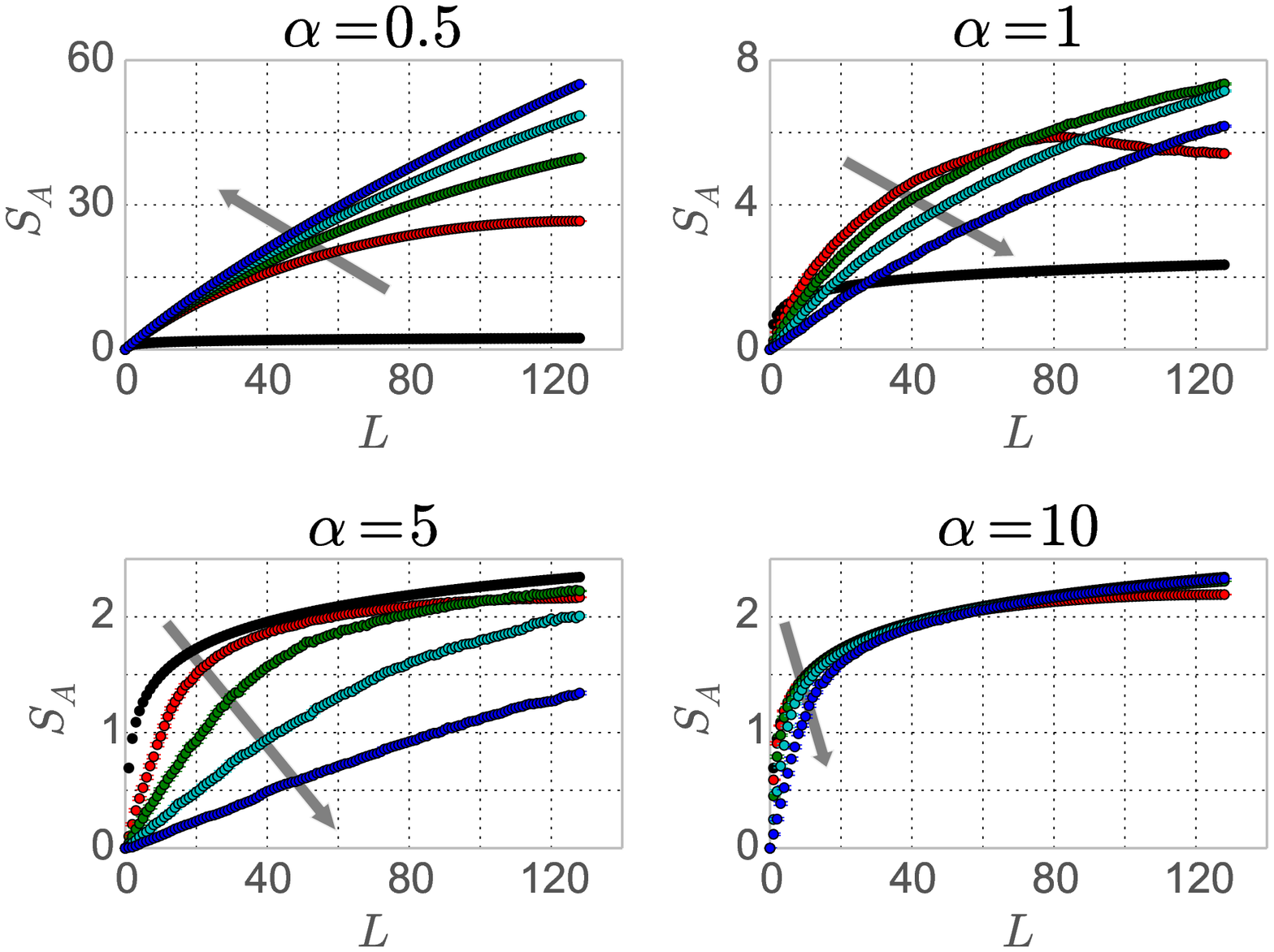}}
\scalebox{0.45}{\includegraphics{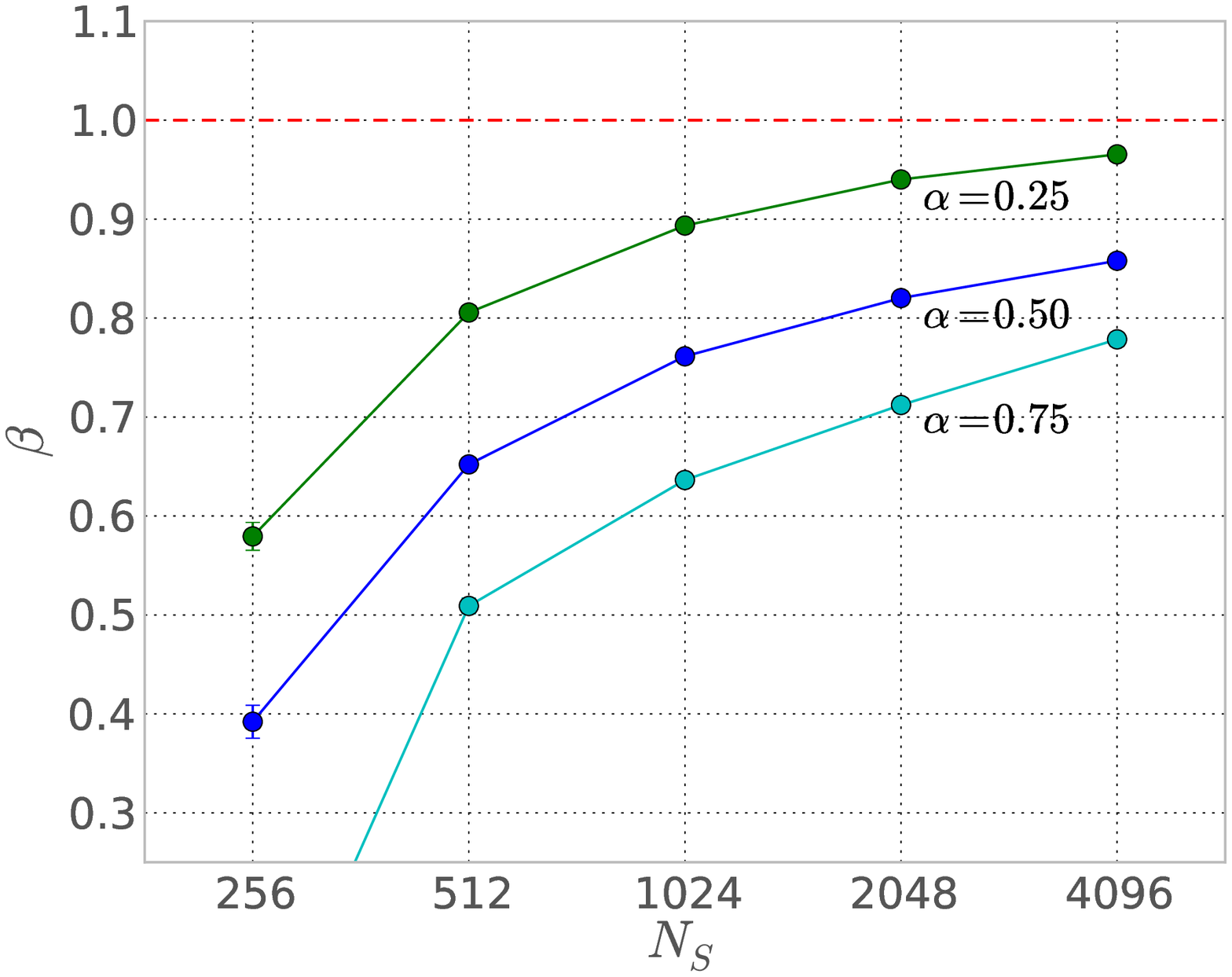}}
}
\caption{Left panel: entanglement entropy calculated for four values
of the decay exponent $\alpha=0.5$,$1$,$5$,$10$ in the
random models (\ref{random}) for the sizes $N_S=256, 512, 1024, 2048,4096$ (and 
half-filling): the curves are obtained averaging 
over $400$ realization of the disorder.
The gray arrow indicates that the curves have increasing
values of $N_S$. The curve plotted in filled black circles
represent the EE for a short-range system used as a reference 
(note the different scales of the $y$-axes). Right panel: value of the fitted
exponent $\beta$ for three different values of $\alpha$ - from top 
to bottom $\alpha=0.25, 0.5, 0.75$.}
\label{fig10}
\end{figure}

Our findings are summarized in Fig. \ref{fig10} and are the following: for 
$\alpha \gg 1$ , the logarithmic behavior of the random short-range model is 
recovered \cite{ramirez14}. When $\alpha$ decreases, at fixed size $N_S$, the EE 
clearly drops off and then, at a value of $\alpha$ of order $1$,  the EE grows back again, 
as shown in the left part of Fig. \ref{fig10}. We observe that, for $\alpha \ll 1$, the EE 
is larger than the one of the random short-range model and it appears to be approximately 
linear. 

To look for the asymptotic behavior of $S_A$, it is most convenient to allow $L$ 
to vary in a fixed interval (for example between $1$ and $128$) and vary $N_S$. It turns out 
that, for  $\alpha \lesssim 1$, a reasonable fit function has the 
form $S_A(L)=a+b L^\beta$. Varying $N_S$ we plot, in the right part of 
Fig. \ref{fig10}, $\beta$ as a function of $N_S$ for three different values 
of $\alpha<1$. For $\alpha \gtrsim 1$ the EE decreases as $N_S$  
increases and there is  an interval 
of values of $\alpha$ where the EE 
becomes even smaller than the one of the clean short-range model \cite{note1}. 
For $\alpha < 1$, one  finds a power-law behavior of $S$, and $\beta$ appears to grow as $N_S$ increases. 

Even if not conclusive, these results seem to indicate that another way to generate
a violation of the area law for the EE of the ground state is due to the effect of disorder.  

\begin{thebibliography}{XX}
\bibitem{note}
Indeed, considering the dimension of $\mathcal{B}$ beyond 
$\dim \mathcal{B} > \dim \bar{\mathcal{A}}$,
proves to be detrimental for getting higher EE states.
In fact if we switch $\mathcal{B}$ and $\bar{\mathcal{B}}$ due to particle-hole
simmetry we see that the number of nonzero eigenvalues
of $C$ gets smaller than $\dim \mathcal{A}$.
\end{thebibliography}
\end{document}